\documentclass{aa}  
\usepackage{graphicx}
\usepackage{array}
\usepackage{tabularx}
\usepackage{natbib}
\setcitestyle{aysep={}} 
\bibpunct{(}{)}{;}{a}{}{,} 
\usepackage[draft=False, colorlinks=true, linkcolor=blue, urlcolor=blue, citecolor=blue]{hyperref}
\usepackage{longtable}
\usepackage{lscape}
\usepackage{siunitx}
\usepackage{color,ulem}
\usepackage{nicefrac}
\usepackage[T1]{fontenc}
\usepackage{txfonts}
\usepackage{xspace}

\usepackage{booktabs}  

\newcommand{\mj}{\ensuremath{M_\mathrm{Jup}}\xspace}
\newcommand{\rj}{\ensuremath{R_\mathrm{Jup}}\xspace}
\newcommand{\mjyr}{\ensuremath{M_\mathrm{Jup}\,\mathrm{yr^{-1}}}\xspace}
\newcommand{\Msyr}{\ensuremath{M_\odot\,\mathrm{yr^{-1}}}\xspace}
\newcommand{\lsun}{\ensuremath{L_\odot}\xspace}
\newcommand{\msun}{\ensuremath{M_\odot}\xspace}

\newcommand{\Mp}{\ensuremath{M_{\mathrm{p}}}\xspace}
\newcommand{\Rp}{\ensuremath{R_{\mathrm{p}}}\xspace}
\newcommand{\Mstar}{\ensuremath{M_\star}\xspace}

\newcommand{\Teff}{\ensuremath{T_{\mathrm{eff}}}\xspace}
\newcommand{\Ha}{\ensuremath{\mathrm{H}\alpha}\xspace}
\newcommand{\Hb}{\ensuremath{\mathrm{H}\beta}\xspace}

\newcommand{\Lbol}{\ensuremath{L_{\mathrm{bol}}}\xspace}
\newcommand{\Lacc}{\ensuremath{L_{\mathrm{acc}}}\xspace}
\newcommand{\mdot}{\ensuremath{\dot{M}_{\mathrm{acc}}}\xspace}
\newcommand{\logm}{\ensuremath{\mathrm{log}(\mdot/\mjyr)}\xspace}
\newcommand{\Lline}{\ensuremath{L_{\mathrm{line}}}\xspace}
\newcommand{\logla}{\ensuremath{\mathrm{log}(\Lacc/\lsun)}\xspace}
\newcommand{\logll}{\ensuremath{\mathrm{log}(\Lline/\lsun)}\xspace}
\newcommand{\Hi}{H~\textsc{i}\xspace}
\newcommand{\Hei}{He~\textsc{i}\xspace}
\newcommand{\ff}{\ensuremath{f_{f}}\xspace}
\newcommand{\lamrest}{\ensuremath{\lambda_\mathrm{rest,\,air}}\xspace}
\newcommand{\lamobs}{\ensuremath{\lambda_\mathrm{observed}}\xspace}
\newcommand{\Wten}{\ensuremath{W_{10}}\xspace}
\newcommand{\kms}{\ensuremath{\mathrm{km\,s}^{-1}}\xspace}

\newcommand{\fluxinteg}{\ensuremath{\mathrm{erg\,s^{-1}\,cm^{-2}}}\xspace}

\newcommand{\cmcube}{\ensuremath{\mathrm{cm}^{-3}}\xspace}

\makeatletter
\renewcommand*\aa@pageof{, page \thepage{} of \pageref*{LastPage}}
\makeatother

\begin{document} 
   \title{Exoplanet accretion monitoring spectroscopic survey (ENTROPY)}
\subtitle{I. Evidence for magnetospheric accretion in the young isolated planetary-mass object 2MASS J11151597+1937266\thanks{Based on observations collected at the European Southern Observatory under ESO programme 0111.C-0166(A).}}
    \titlerunning{Exoplanet Accretion Monitoring Spectroscopic Survey (ENTROPY). I.}
    \authorrunning{G.\ Viswanath et al.}

   \author{Gayathri~Viswanath
          \inst{\ref{stockholm}}
          \and
          Simon~C.~Ringqvist
          \inst{\ref{stockholm}}
          \and
          Dorian~Demars
          \inst{\ref{grenoble}}
          \and
          Markus~Janson\inst{\ref{stockholm}}\
          \and
          Micka{\"e}l~Bonnefoy\inst{\ref{grenoble}}
          \and
          Yuhiko~Aoyama\inst{\ref{tsinghua_adv}, \ref{tsinghua_astro}, \ref{tokyo}}
          \and
          Gabriel-Dominique~Marleau\inst{\ref{heidelberg}, \ref{duisburg}, \ref{tubingen}, \ref{bern}}
          \and
          Catherine~Dougados\inst{\ref{grenoble}}
          \and
          Judit~Szul\'agyi\inst{\ref{eth}}
          \and 
          Thanawuth~Thanathibodee\inst{\ref{umichigan}, \ref{uboston}}
          }

   \institute{Institutionen f\"{o}r astronomi, Stockholms universitet, AlbaNova universitetscentrum, 106 91,                  Stockholm, Sweden\\
        \email{gayathri.viswanath@astro.su.se}
            \label{stockholm}
            \and
            Univ. Grenoble Alpes, CNRS, IPAG, 38000 Grenoble, France 
            \label{grenoble}
            \and
            Institute for Advanced Study, Tsinghua University, Beijing 100084, PR China 
            \label{tsinghua_adv}
            \and
            Department of Astronomy, Tsinghua University, Beijing 100084, PR China 
            \label{tsinghua_astro}
            \and
            Department of Earth and Planetary Science, The University of Tokyo, 7-3-1 Hongo, Bunkyo-ku, Tokyo 113-0033, Japan 
            \label{tokyo}
            \and
            Max-Planck-Institut f\"ur Astronomie, K\"onigstuhl 17, 69117 Heidelberg, Germany
            \label{heidelberg}
            \and
            Fakult\"at für Physik, Universit\"at Duisburg--Essen, Lotharstra\ss{}e 1, 47057 Duisburg, Germany
            \label{duisburg}
            \and
            Institut f\"ur Astronomie und Astrophysik, Universit\"at T\"ubingen, Auf der Morgenstelle 10, 72076 T\"ubingen, Germany
            \label{tubingen}
            \and
            Physikalisches Institut, Universit\"at Bern, Gesellschaftsstr.~6, 3012 Bern, Switzerland
            \label{bern}
            \and
            ETH Z\"urich, Department of Physics, Wolfgang-Pauli-Str.~27, CH-8093, Z\"urich, Switzerland
            \label{eth}
            \and
            Department of Astronomy, University of Michigan, 323 West Hall, 1085 South University Avenue, Ann Arbor, MI 48109, USA
            \label{umichigan}
            \and
            Institute for Astrophysical Research and Department of Astronomy, Boston University, 725 Commonwealth Ave., Boston, MA 02215, USA
            \label{uboston}
             }

   \date{Received -- ; accepted --}
 
  \abstract
  {Accretion among planetary mass companions is a poorly understood phenomenon, due to the lack of both observational and theoretical studies. Detection of emission lines from accreting gas giants facilitate detailed investigations into this process.}
  {This work presents a detailed analysis of Balmer lines from one of the few known young, planetary-mass objects with observed emission, the isolated L2$\gamma$ dwarf 2MASS J11151597+1937266 with a mass between 7 and~21~\mj and an age of 5--45~Myr, located at $45\pm2$~pc.}
  {We obtained the first high-resolution ($R\sim50,000$) spectrum of the target with VLT/UVES, an echelle spectrograph operating in the near-UV to visible wavelengths (3200--6800~\AA).}
  {We report several resolved \Hi (H3--H6) and \Hei emission lines ($\lambda5875.6$) in the spectrum. Based on the asymmetric line profiles of \Ha and \Hb, 10\% width of \Ha ($199\pm1$~\kms), tentative \Hei~$\lambda6678$ emission and indications of a disk from MIR excess, we confirm ongoing accretion at this object. Using the \textit{Gaia} update of the parallax, we revise its temperature to $1816\pm63$ K and radius to $1.5\pm0.1$ \rj. Analysis of observed \Hi profiles using 1D planet-surface shock model implies a pre-shock gas velocity of $v_0=120^{+80}_{-40}$~\kms and a pre-shock density of $\log(n_0/\mathrm{cm}^{-3})=14^{+0}_{-5}$. The pre-shock velocity points to a mass of $\Mp=6^{+8}_{-4}\,\mj$ for the target. Combining the \Hi line luminosities and planetary \Lline--\Lacc scaling relations, we derive a mass accretion rate of \mdot$=1.4^{+2.8}_{-0.9}\times10^{-8}$~\mjyr.}
  {The line-emitting area predicted from planet-surface shock model is very small ($\sim0.03\%$), and points to a shock at the base of a magnetospherically induced funnel. The \Ha profile exhibits much stronger flux than predicted by the model that best fits the rest of the \Hi profiles, indicating that another mechanism than shock emission contributes to the \Ha emission. Comparison of line fluxes and mass accretion rates from archival moderate-resolution SDSS spectra indicate variable accretion at 2MASS J11151597+1937266.} 
  

  \maketitle
%
\nolinenumbers
\section{Introduction} \label{intro}

The exact nature of accretion among sub-stellar companions has been the subject of several studies for the last two decades. In the stellar regime, the accretion process is well observed and understood, with the collapse of a molecular cloud resulting in a protostar and a circumstellar disk \citep{adams1987}. The protostar then grows over the next few million years by accreting material from the inner regions of the disk (separated from the star due to the strong magnetic field), along its magnetic field lines, in a process called magnetospheric accretion \citep{koenigl1991, hartmann1998}. Whether the same formation process extends to the sub-stellar regime is still unclear. An alternative theory is that planets have insufficient magnetic field strengths to truncate their disks and instead accrete material via a boundary layer \citep{lynden1974, owen2016}; recent studies have however indicated that newly formed giant planets and brown dwarfs can possess magnetic field strengths of up to a few kilogauss \citep{reiners2010, batygin2018, kao2018}. Consequently, there is ambiguity regarding whether hydrogen emission lines from accreting giant protoplanets originate from the gas in magnetospheric accretion funnels \citep{thanat2019}, similar to the stellar case \citep{calvet1998, muzerolle2001}, or from shock-heated gas at the surface of the planet or from the circumplanetary disk (CPD), or from both \citep{szulagyi2014, zhu2015, szulagyi2017, aoyama2018, szulagyi2020, aoyama2021}. It is also yet to be confirmed if the decreasing trend seen among stars and brown dwarfs between their accretion rate and mass \citep{rebull2000, muzerolle2003, natta2004, mohanty2005, muzerolle2005, venuti2019} extends down to the planetary mass range. A major reason behind the lack of a conclusive explanation on formation of brown dwarfs and planetary mass companions (PMCs) has been the absence of observational evidence of accretion from such objects, mainly due to the technical limitations up to now. 

With the recent era of sensitive, high-resolution instruments, there has been a rising number of observations of accreting planetary mass objects in the last few years. The PDS 70 system serves as a benchmark in this context, with H$\alpha$ detection from both its protoplanets b and c (in $\sim20$ and $\sim30$~au orbits respectively) at several epochs via circumstellar disk imaging \citep{wagner2018, haffert2019, hashimoto2020, zhou2021} and ALMA \citep{alma} detection of the CPD around PDS 70c \citep{isella2019, benistry2021}. H$\alpha$ emission has also been reported around the protoplanetary candidate LkCa 15b \citep{sallum2015}; however imaging and spectroscopic follow-up at subsequent epochs \citep{whelan2015, mendi2018, currie2019, blakely2022} have attributed the detection to disk features and extended H$\alpha$ emission around the star LkCa 15. 
Lately, accretion signatures have been detected spectroscopically in the form of \Hi emission lines from brown dwarf and planetary mass companions like Delorme 1 (AB)b \citep{erikson2020, betti2022, ringqvist2023}, TWA 27B \citep{luhman2023}, GSC 06214-00210 b, GQ Lup b \citep{demars2023} and SR 12C \citep{sr12c2018,sr12c2018erratum}, which are all essentially isolated companions in wide orbits ($\gtrsim50$~au). High resolution observations of emission lines from such accreting targets enable detailed investigations of line luminosities and profile asymmetries, which in turn give a wealth of information about the nature of accretion process itself. Such direct spectroscopic observations of low-mass objects in the cradle of formation help in the continuing effort to constrain formation mechanisms in the sub-stellar regime and understand the differences in accretion process (if any) not only among planets, brown dwarfs and stars but also between planetary mass objects in close orbits (like the PDS 70 system) and isolated/essentially isolated environments.

The target of this study, 2MASS J11151597+1937266 (hereafter 2M1115), joins the slow-growing sample accreting sub-stellar objects that can be directly observed. It was identified by \cite{theissen2017} as a young, low surface gravity object of spectral type L2 as part of the LaTE-MoVeRS survey. The object showed strong hydrogen (\Hi) and helium (\Hei) emission in its optical spectrum, as well as mid-infrared (MIR) excess in its spectral energy distribution (SED). A follow-up study by \cite{theissen2018} points to the presence of \Hi and \Hei emission lines in the moderate-resolution (R$\sim$2000) optical spectrum from multiple SDSS \citep[Sloan Digital Sky Survey;][]{sdss} epochs, indicating that the object possesses either persistent enhanced magnetic activity or weak accretion or both. Based on the signs of possible ongoing accretion and the indications of a disk around the object from the excess MIR flux, the age was loosely constrained to 5--45 Myr. Using its low-resolution (R$\sim$120) near-infrared (NIR) spectrum from SpeX \citep{spex}, the spectral type of the object was constrained to L2$\gamma$ and the mass to 7--21 \mj~based on evolutionary models. Table~\ref{tab1} summarises the known properties of 2M1115, including the updated distance information from \textit{Gaia} \citep[object ID: Gaia DR3 3990192705824438144;][]{gaiadr3}. \cite{theissen2018} also identified a potential co-moving, young ($\lesssim$100 Myr) M-type star 2MASS J11131089+2110086 (hereafter 2M1113) at an angular separation of 1.62$^{\circ}$ from the target with similar proper motion and radial velocities. Neither 2M1115 nor 2M1113 has been identified to be associated to any known nearby young moving groups (NYMGs), posing the possibility of being members of a yet undiscovered kinematic group or being ejected by a known NYMG. 2M1115 thus joins the growing population of young, isolated low mass stellar and sub-stellar objects, and is also relatively nearby ($45.21\pm2.20$~pc). This makes the object an ideal target to study sub-stellar formation scenarios and pave the way to a deeper understanding of the nature of accretion in brown dwarfs and PMCs.

Here, we present the first high resolution observations of 2M1115 in the optical to near-UV wavelength range (described in Section \ref{obs}). The data reduction is described in Section \ref{reduct}. In Section \ref{results}, we outline the results from the observations, with the detection of several emission lines that offer an in-depth look into the accretion at this young PMC. A detailed investigation of the line profiles is also presented in this section, along with a discussion on the possible association with 2M1113. Section \ref{disc} outlines the implication from these results.

\renewcommand{\arraystretch}{1.2}
\begin{table}[!ht]
\caption{Basic parameters for 2M1115 from the literature and this work. }
\centering
\begin{tabular}[l]{l l l}
\hline\hline
\textrm{Parameter} & \textrm{Value}  & \textrm{Reference} \\\hline
& Measured & \\\hline
ICRS RA ($\alpha$, ep=2016.0) &  $168.816234^{\circ}$ &  1 \\
ICRS Dec.\ ($\delta$, ep=2016.0)  &
$+19.623939^{\circ}$  & 1  \\
Parallax [mas] & $22.12 \pm 1.07$ & 1 \\
$\mu_{\alpha*}=\mu_\alpha \cos(\delta)$ [mas\,yr$^{-1}$] & $-69.53\pm 1.32$ & 1 \\
$\mu_{\delta}$ [mas\,yr$^{-1}$] & $-25.21\pm 1.48$ & 1 \\
G [mag] & $20.41\pm0.02$ & 1 \\
J [mag] & $15.59\pm0.06$  & 2 \\
H [mag] & $14.57\pm0.06$ & 2 \\
$K_S$ [mag] & $13.80\pm0.05$ & 2 \\
W1 (ALLWISE) [mag] & $13.09\pm0.02$ & 3 \\
W2 (ALLWISE) [mag] & $12.55\pm0.03$ & 3 \\
W3 (ALLWISE) [mag] & $10.77\pm0.12$ & 3 \\
W4 (ALLWISE) [mag] & $>8.68$ & 3 \\\hline
 & Derived & \\\hline
Distance [pc]  & $45.21\pm2.20$ & 1 \\
Spectral Type & L2$\gamma\pm1$ & 4 \\
RV [km\,s$^{-1}$] & $-14\pm 7$ & 4 \\
Age [Myr] & 5--45 & 4 \\
Mass [$M_\mathrm{Jup}$] & 7--21\tablefootmark{a} & 4 \\
 & $6^{+8}_{-4}$ & 5 \\
Radius [$R_\mathrm{Jup}$] & $1.5\pm0.1$ & 5 \\
\Teff~[K] & $1816\pm63$ & 5 \\
$\log(\Lbol/\lsun)$ & $-3.63\pm0.05$ & 5 \\\hline
\end{tabular}
\tablefoot{\\ \tablefoottext{a}{The mass estimate from \cite{theissen2018} depends on their photometric distance estimate for the target, $d_\mathrm{phot}=37\pm6$~pc (but see Section~\ref{disc}).} }
\tablebib{
(1)~\citet{gaiadr3}; (2)~\citet{cutri2003}; (3)~\citet{wise2014}; (4)~\citet{theissen2018}; (5)~This work
}
\label{tab1}
\end{table}

\section{Observations with VLT/UVES} \label{obs}
2M1115 was observed as part of the ``ExoplaNeT accRetion mOnitoring sPectroscopic surveY'' (ENTROPY) survey with the Ultraviolet and Visual Echelle Spectrograph \citep[UVES;][]{uves} mounted at Unit Telescope~2 of the Very Large Telescope (VLT) at the Paranal Observatory, Chile, between 2023 June 10--11 (MJD 60105, 60106). A total of four frames at 740~s exposure each were obtained across the two nights, giving a total integration time of 0.82~hr. The seeing was measured to be $\sim1.43\arcsec$ on an average across both nights of observations, at an average air mass of $\sim1.416$. The observing log is given in Table~\ref{aptab1}. The observations were carried out in Dichroic \#1 mode with both the blue and red arms of the instrument, centred at 390~nm and 580~nm respectively, with overlapping wavelength ranges in the subsequent orders of each arm. Between the two arms, and using a slit width of 0.8$\arcsec$, these observations offer a high resolution\footnote{\url{https://www.eso.org/observing/dfo/quality/UVES/reports/HEALTH/trend_report_ECH_RESOLUTION_DHC_HC.html}} of $R_{\lambda}\sim$50,000--53,000 over wavelengths at 3200--6800~\AA. 

\section{Data reduction} \label{reduct}
Since the continuum of the target is very weak in the observed wavelength range, only its emission lines are readily visible in the raw data. As a result, the automatic ESO/UVES reduction and extraction pipeline failed to produce usable results. We therefore implemented a manual data reduction scheme, based on the raw data and existing calibration files. Basic reduction of the data was performed by subtracting the bias, correcting for flat-field and removing the inter-order background. Cosmic rays were accounted for by using a horizontal median filtering, as well as by masking out pixels higher than 10~times the photon noise at the location. Extraction of the object spectrum was performed using standard aperture photometry within an aperture size of 10 pixels in the spatial direction for the blue arm and 20~pixels for the red arm, centred on the object location in the slit. Sky background and telluric lines were removed by subtracting the flux within the respective aperture sizes of the blue and red arms, but with the apertures placed instead at both the ends of the slit. Wavelength calibration was performed using the arc lamp spectrum and a Th--Ar line list.

The 1D spectrum thus extracted was flux calibrated based on the instrument's response curve derived from the observations of a standard star taken with the same observing set-up as the target  (see Section~\ref{apendixb} for further details).
Since the seeing was greater than 1$\arcsec$ for all four observations, the calibrated flux was compensated for the relative slit-loss from seeing ($\sim4$\%) between the observations of the target and that of the standard star, by determining a 1D point spread function (PSF) along the spatial direction in the data, constructing a corresponding 2D PSF by assuming circular symmetry, and calculating the fractional flux lost outside of the 0.8$\arcsec$ wide slit in the dispersion direction. A barycentric velocity correction was applied to the wavelength calibration for each frame using {\fontfamily{qcr}\selectfont
astropy.skycoord} at Cerro Paranal at the respective time of observation. A stacked spectrum was then obtained by using the weighted average across all four observations, with the weights determined from the respective photon noise. The flux uncertainties were determined by the weighted standard deviation across the individual spectra. The stacked, calibrated UVES spectrum is available through ESO archive.

\section{Results and Analysis} \label{results}
We report the first resolved \Hi and \Hei emission line detections from 2M1115 in the optical to near-UV wavelengths. We present confirmed detection of the \Hi emission lines \Ha to H6, and \Hei line at $\lambda5875.62$\footnote{In this work, all wavelengths have been taken from the 
NIST Atomic Spectra Database, available at \url{https://www.nist.gov/pml/atomic-spectra-database}. The wavelengths are in air.} (listed in Tables~\ref{aptab2} and~\ref{aptab3}). Tentative detections of H7, \Hei~$\lambda6678.15$ and possible metal lines Ca~\textsc{ii}~H $\lambda 3968.47$, Fe~\textsc{i} $\lambda\lambda 3834.22$, 3967.42, 4839.54, 4970.50, 4986.22, 5329.99, 5684.43, 6083.66, 6636.96, and 6703.57, Ti~\textsc{i} $\lambda5720.43$ and Cr~\textsc{i} $\lambda5991.07$ have also been noted but are not statistically significant with the signal-to-noise ratio (S/N) of the existing data; these are listed in Table~\ref{aptab3}. The identification scheme for the confirmed and tentative lines are described in Appendix~\ref{apendixc}.

\subsection{Characteristics of neutral-hydrogen lines} \label{profilechar}
We detected \Hi emission lines from H3--H6 with strong S/N ($>5\sigma$) in the spectrum. H4 (\Hb) was detected in two consecutive orders 2 and 3 of the lower Red (RedL) arm, but had an S/N of only $4.3\sigma$ in order 3 as it was near the edge of the order where the spectrum was generally noisier. Hence, we only include the stronger detection in order 2 ($8.7\sigma$) in our main analysis. H5 (H$\gamma$) was also detected in orders 34 and 35 of the Blue arm with similar S/N ($4.6\sigma$), resulting in a root mean square (rms) S/N of $6.5\sigma$; hence the average of their respective line fluxes is used for the main analysis. Although H7 was detected in both orders 24 ($1.7\sigma$) and 25 ($3.5\sigma$) of the Blue arm, the rms S/N was still below $5\sigma$ and is thus classified as a tentative detection. 

\begin{figure}[ht]
\centering
\includegraphics[width=1.0\linewidth, trim = {1cm 0cm 2cm 0.5cm}, clip]{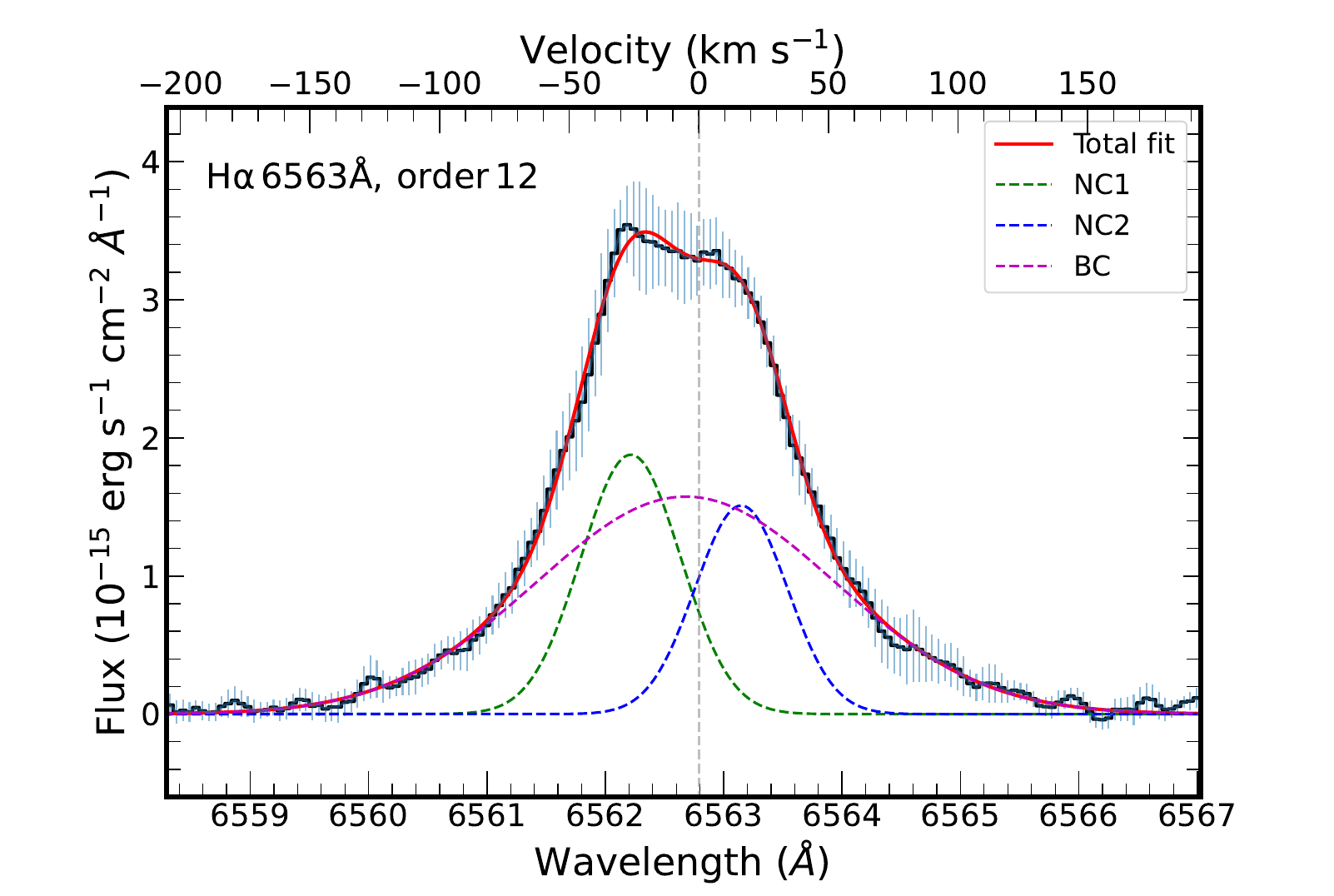} \\
\includegraphics[width=1.0\linewidth, trim = {1cm 0cm 2cm 0.5cm}, clip]{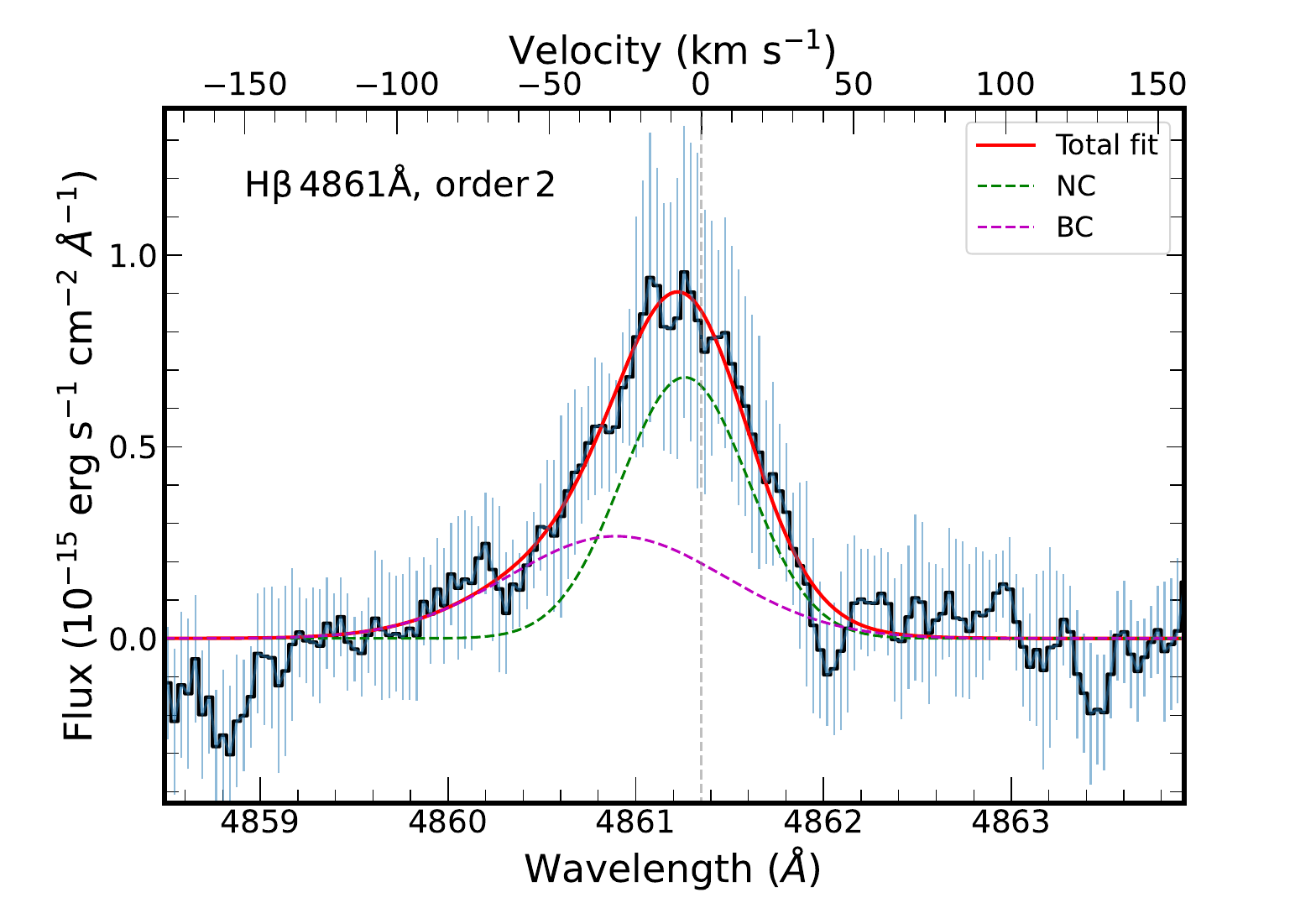}
\caption{Line profiles of \Ha in the RedU arm (top panel), and \Hb in the RedL arm (bottom panel) of the UVES spectrum of 2M1115. NC, BC indicate narrow component and broad component, respectively, of the total Gaussian fit to the profile. The flux uncertainties are shown as vertical error bars and represent the weighted standard deviation among the four observations. Velocity is set with respect to the rest-frame wavelength (in air) of the respective lines.}
\label{fig1}
\end{figure}

\begin{figure}[ht]
\centering
\includegraphics[width=0.5\textwidth]{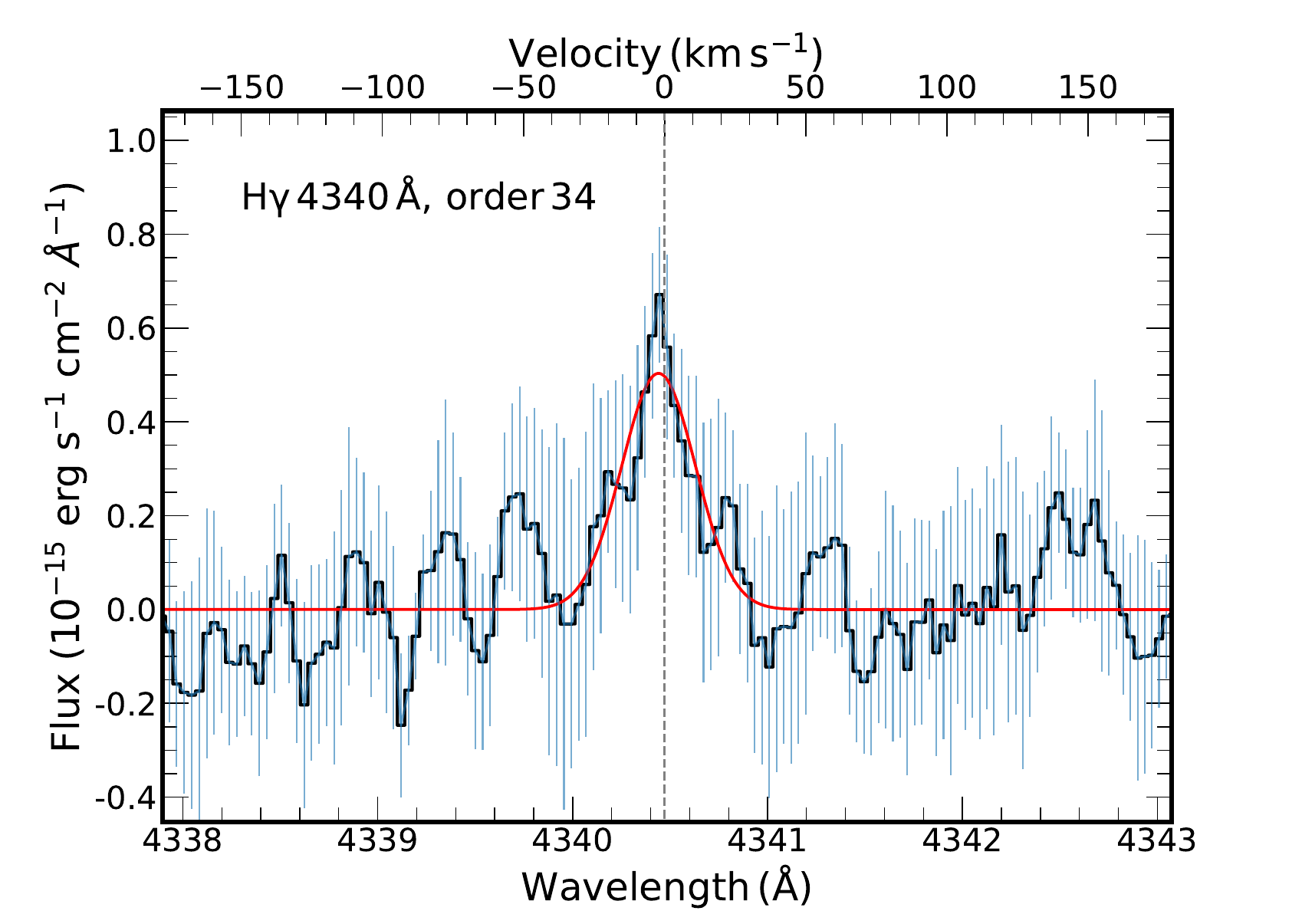} \\
\includegraphics[width=0.5\textwidth]{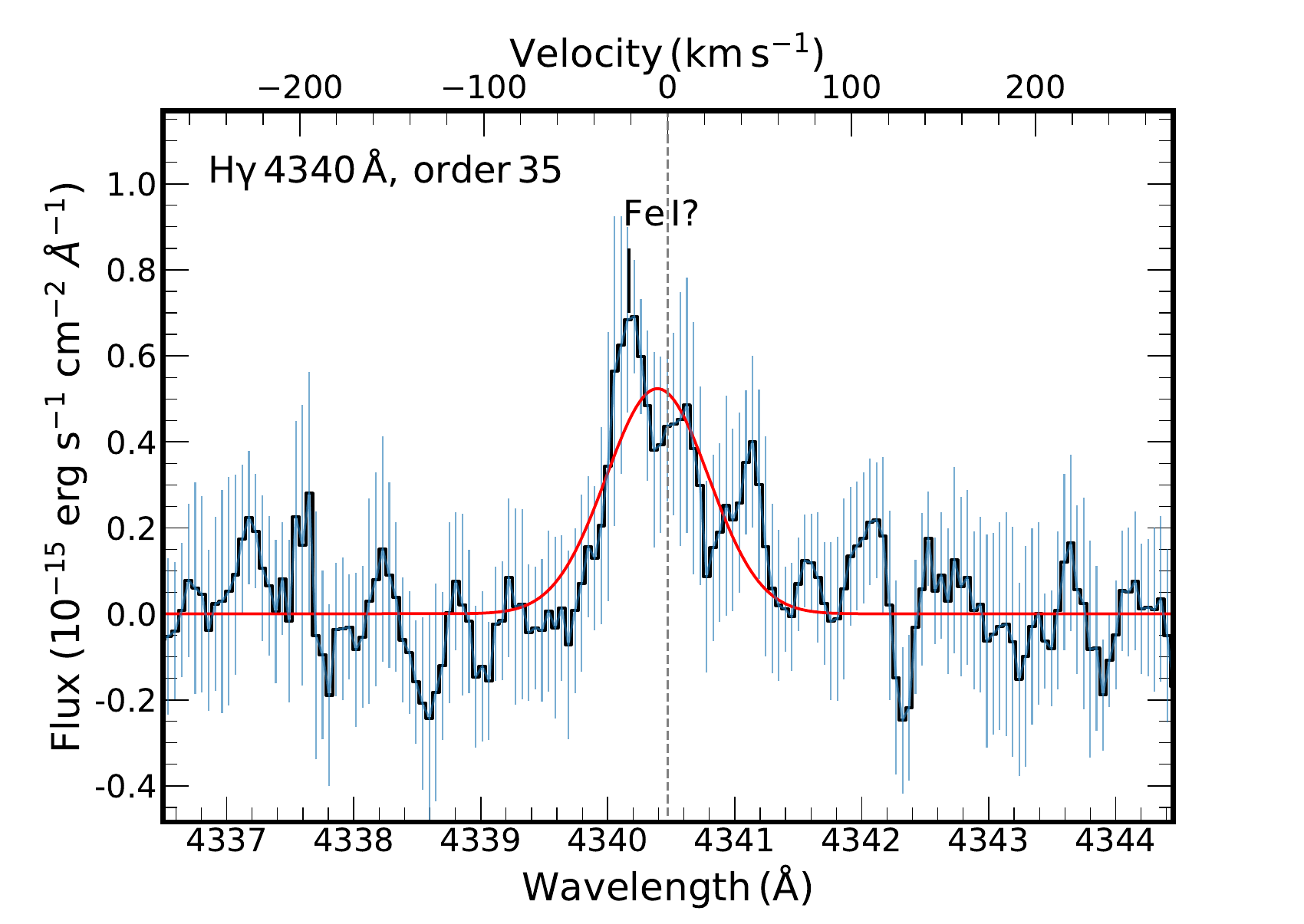}\\
\includegraphics[width=0.5\textwidth]{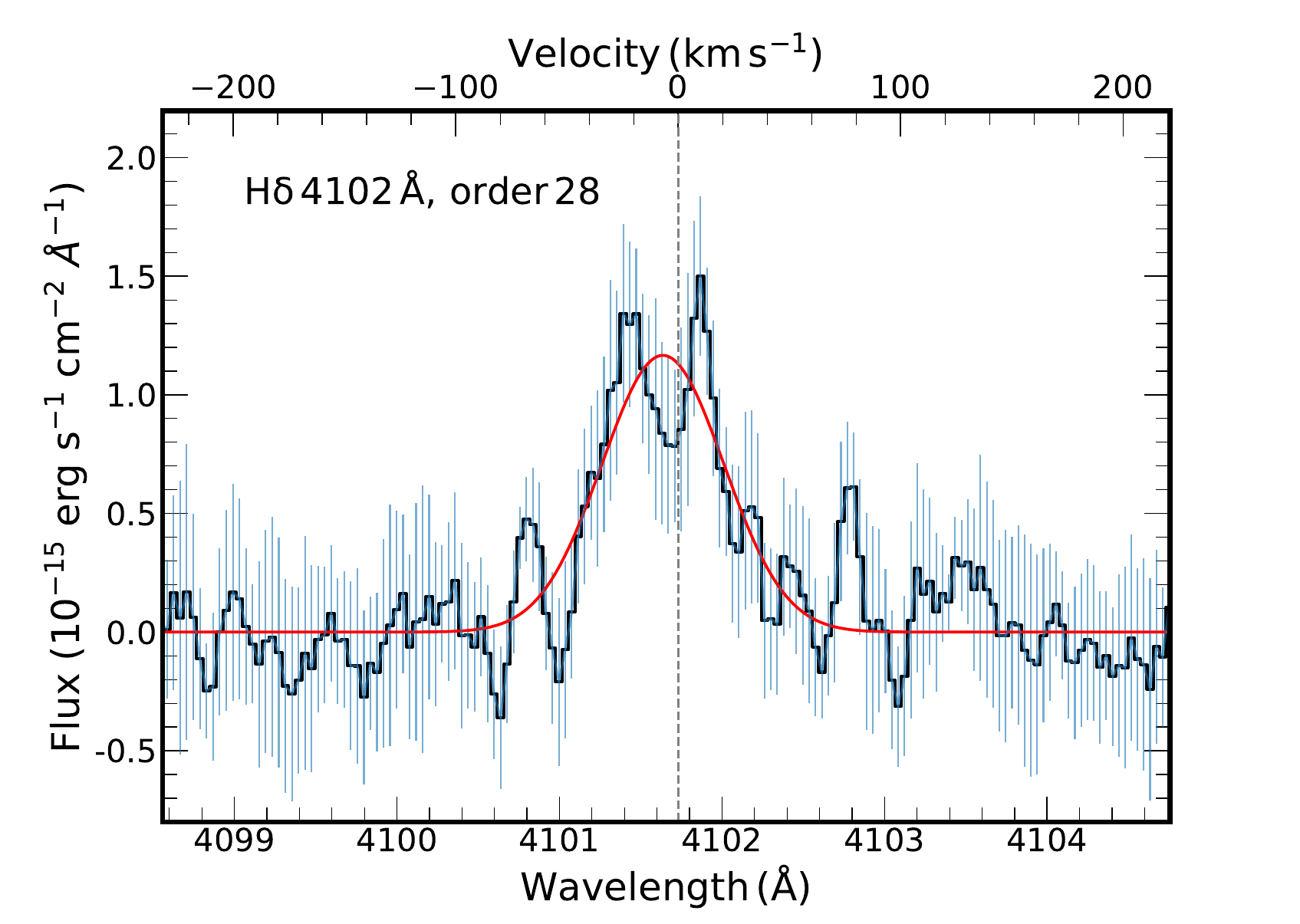}
\caption{Line profiles of (upper panel) H$\gamma$ in order 34, (middle panel) H$\gamma$ in order 35, and (lower panel) H$\delta$ in order 28 detected in the UVES spectrum for 2M1115 in the Blue arm. The flux uncertainties are shown as vertical error bars and represent the weighted standard deviation among the four observations. Velocity is set with respect to the rest-frame wavelength (in air) of the respective lines.}
\label{fig2}
\end{figure}

The line profiles were determined by performing Gaussian fits on the average-smoothed spectrum with a box size of 3 pixels, and are shown in Figure~\ref{fig1} and Figure~\ref{fig2} for the confirmed line detections, H$\alpha$--H6. The integrated line fluxes and luminosities of H~\textsc{i} lines estimated thus are listed in Table~\ref{tab2} (refer Table~\ref{aptab2} for profile characteristics). We note that H6 is brighter than H5 in the spectrum\footnote{Similar behaviour has also been observed in the $\sim40$ Myr PMC Delorme~1~(AB)b, where H6 was observed to be overluminous \citep[see ][]{ringqvist2023}.} and is double peaked in all four individual spectra (see Figure~\ref{fig2}, lower panel). The increase in the line flux corresponding to the two peaks could be caused by possible blending with Fe~\textsc{i} lines at 4101.26~\AA\ and 4101.65~\AA. H5 in order 35 of the Blue arm has a clear flux increase at the bluer end of its line profile (see Figure~\ref{fig2}, upper panel). We suspect the cause to be possible blending with a metal line; most likely with Fe~\textsc{i} at 4340.49~\AA. However the S/N is not high enough to make statistically reliable claims based on these features.

\subsection{Asymmetry in the line profiles} \label{assymetry}

Among the detected \Hi lines, \Ha and \Hb show clear asymmetries in their line profiles. Following \cite{ringqvist2023}, we fit multi-component Gaussian models to their line profiles, composed of a combination of narrow (NC) and broad (BC) components. The best model for each line was determined as the one that gave the least $\chi^2$, calculated as
\begin{equation}
    \chi^2 = \sum_{i} \left[ \frac{N_i - fi}{\sigma_i} \right]^2\,,
\end{equation}
where $\sigma_i$ is the error in the measurement $N_i$, and $f_i$ is the corresponding model prediction. The resulting line profile fits for \Ha and \Hb are given in Figure~\ref{fig1} and their characteristics are given in Table~\ref{aptab2}. Such a multi-component profile fit is reasonable as explained in the 1D planet-surface shock model of accretion in \cite{aoyama2018, aoyama2020, aoyama2021}, where the hotter ($\gtrsim10^5$~K) gas in the immediate post-shock region is receding away from the observer into the planet and leads to a broad, redshifted profile. Comparatively, gas coming from deeper below the shock region originates at lower temperatures ($\approx10^4$~K) and is thus narrower. In Section~\ref{disc}, we discuss the significance of these profile fits further.

We also fit the obtained \Hi line profiles with predictions from \cite{aoyama2020, aoyama2021} models as a function of pre-shock velocity $v_0$ and pre-shock number density $n_0$. We allowed the model grid to vary from $v_0= 50-200~\kms$ and $n_0=10^9-10^{14}~\cmcube$ during the fitting process. The observed \Ha emission from the target is quite strong compared to the other \Hi lines. One reason could be the strong dependence of extinction on wavelength, as seen in PDS 70b and c \citep{hashimoto2020}. On setting the extinction to a high value of $\mathrm{A_v=3.5}$~mag, the model shows a good fit to both \Ha and \Hb, but very weak predictions for H6 and H7, making the overall model prediction unsatisfactory. A simple interpretation from this analysis could be that the main origin for \Ha emission is not the shock-heated region. Aoyama et al.\ (in prep.) have demonstrated similar, strong contamination for \Ha with other than shock-heated gas (e.g., gas along accretion columns or chromospheric activity) using archival data for massive planets $>10$~\mj. In these cases the line profiles show better agreement with the model for \Hb and higher order lines, on excluding \Ha from the analysis. Accordingly, we repeated our fitting process by including only the lines H4--H7 and fixing the extinction to $\mathrm{A_v=0}$~mag. The resulting fit reproduces the H4--H7 line profiles very well within their flux uncertainties, with model parameters of $v_0=120^{+80}_{-40}~\kms$ and $\log(n_0/\mathrm{cm}^{-3})=14^{+0}_{-5}$ (see Section~\ref{shockmodel} for details on the fitting process and the inference of uncertainties on these parameters). This strongly suggests that shock emission contributes only minorly to \Ha. On the other hand, the higher-order Balmer lines H4--H7 are well-reproduced with the shock emission model in this analysis. As seen from the observed hydrogen line spectral profiles in T-Tauri stars, higher-order Balmer lines are generally less contaminated by emission from accretion columns since the lower temperature of the accretion flow at these low mass stars allow only the lower-order hydrogen lines, such as \Ha, to be bright \citep[see][]{hartmann2016}. A similar trend can be expected in planetary mass objects as well. Due to these reasons, there is less likelihood for the detected H4--H7 lines to have contribution from the accretion columns. In Section \ref{disc}, we show that the expected level of chromospheric emission based on the target's temperature is not significant compared to the observed Balmer emission. However, since chromospheric activity can also be non-thermal, a contamination from this source cannot be completely ruled out with the current data.

Considering the pre-shock gas velocity from the best-fit \cite{aoyama2020, aoyama2021} model as the free-fall velocity of the in-falling gas, we can derive a mass estimate \Mp for the accreting PMC \citep{aoyama2019} using 
\begin{equation}
    \Mp = \frac{ \Rp v_0^2}{2G}.
\end{equation}

The existing radius estimate for 2M1115 in the literature \citep[$\Rp=1.3\pm0.2~\rj$; ][]{theissen2018} is based on an earlier photometric distance measurement of $37\pm6$~pc from before the \textit{Gaia} DR3 observation of the target. With the more accurate \textit{Gaia}-parallax-based distance measurement for the target ($45.21\pm2.20$~pc), the radius estimate was in need of revision. For this purpose, we performed an SED fit to the existing photometry of 2M1115 using BT-Settl AGSS 20009 \citep{allard2012, allard2013, asplund2009}, AMES-Dusty \citep{chabrier2000, allard2001} and DRIFT-PHOENIX \citep{phoenix2003, phoenix2008, phoenix2011} model photospheres via the Virtual Observatory SED analyzer \citep[VOSA\footnote{\url{http://svo2.cab.inta-csic.es/theory/vosa/}};][]{bayo2008} version~7.5 (see Section \ref{sed} for details). The corresponding parameters for the target from the best-fitting models are $\Teff=1816\pm63$~K,  $\log(\Lbol/\lsun)=-3.63\pm0.05$ and $\Rp=1.5\pm0.1~\rj$.
With this radius estimate and $v_0=120^{+80}_{-40}$~\kms, we get a mass estimate of \Mp=$6^{+8}_{-4}$~\mj for 2M1115. The resulting $1\sigma$ mass range (2--14~\mj) implies a lower mass for the object than in the literature (see Table~\ref{tab1}). 

\subsection{Mass accretion rate for 2M1115} \label{mar}

The H~\textsc{i} line luminosities can be used to estimate accretion luminosities for 2M1115 using scaling relations of the form
\begin{equation}
\log(\Lacc/L_{\odot}) = a\times \log(\Lline/L_{\odot}) + b, \label{eq1}
\end{equation}
where $a$ and $b$ represent the fit coefficients for the scaling relations of each transition. Given the existing estimate of mass range for the target ($\Mp\approx7$--21~\mj), we used \Lline--\Lacc relations from \cite{aoyama2021} for the lines H3--H7, formulated for planetary mass objects ($\Mp\approx2$--20~\mj) based on spectrally resolved, non-equilibrium models for hydrogen emission due to a shock on the planetary surface. The resulting accretion luminosities for the target are listed in Table~\ref{tab2}.

We derived mass accretion rates for 2M1115 from the estimated accretion luminosities of the H~\textsc{i} lines based on the established relation for stars
\begin{equation}
\label{eq:Mdot}
\mdot = \left( 1-\frac{\Rp}{R_{\mathrm{in}}} \right)^{-1} \frac{\Lacc\Rp}{G\Mp},
\end{equation}
where \Mp represents the mass of the accreting object, \Rp its radius and $R_{\mathrm{in}}$ the inner radius of a truncated circumplanetary disk in a magnetospheric accretion scenario \citep{hartmann2016}. On setting $R_{\mathrm{in}}=5~\Rp$ \citep{gullbring1998, alcala2017, ringqvist2023}, $\mdot\approx \Lacc\Rp/G\Mp$, which remains valid for a planet-surface shock scenario under the assumption that the entire kinetic energy from the shock is converted into radiation \citep{marleau2019}.

With \Rp=$1.5\pm0.1$~\rj (this work) and \Mp=$14\pm7~\mj$ \citep{theissen2018}, we derived a mass accretion rate for 2M1115 from each detected H~\textsc{i} line (H3--H7) using Equation~\ref{eq:Mdot} (listed in Table~\ref{tab2}). The average mass accretion rate thus obtained is $\log(\mdot/\mjyr)=-7.86\pm0.48$ for the target. In Section \ref{assymetry}, we provide evidence for \Ha possessing significant contribution from non-shock emission, with the likely sources being accretion columns or chromospheric activity. Although the expected emission from the latter at the temperature of 2M1115 is lower by several orders of magnitude than the level of emission seen in these observations (see Section \ref{disc}), any minor contribution from chromospheric activity to the detected \Ha emission can not be completely ruled out. 
Usage of \Lline--\Lacc scaling laws which assume that the entire line luminosity contributes to  accretion luminosity should hence be proceeded with caution in this case. However, since the current observational data on the target limits a quantification of this non-shock contribution, or further analysis of its nature, we have considered the total \Ha luminosity from the target while using scaling laws to estimate its accretion luminosity, and subsequently its mass accretion rate, in this work. The \cite{aoyama2021} scaling laws also come with the associated model-dependent uncertainties which would result in systematic uncertainties in the subsequent calculation of accretion luminosity and mass accretion rates. 

In Appendix~\ref{sdss}, we estimated H3--H7 line fluxes for 2M1115 from its existing, moderate-resolution SDSS DR9 \citep{sdss9} and DR12 \citep{sdss12} spectra and, subsequently, derived mass accretion rates $\log(\mdot/\mjyr)$ of $-8.00\pm0.32$ and $-7.62\pm0.32$ from DR9 and DR12 respectively. The mass accretion rate estimate from the UVES data is, respectively, $\sim 0.3\sigma$ higher and $\sim 0.5\sigma$ lower (in dex) than those from SDSS DR9 and DR12. We discuss this difference further in Section~\ref{disc}.  

Together with the mass accretion rate and the best-fit pre-shock velocity and number density from the analysis within the planet-surface shock framework (Section \ref{assymetry}), we can estimate the fraction of the planetary surface from where the \Hi line radiation is emitted, the filling factor \ff \citep{aoyama2019}, as
\begin{equation}
    \dot{M} = 4\pi \Rp^2 f_{f} \mu^{\prime} n_0v_0.
\end{equation}
Here, $\mu^{\prime}$ is the mean weight per hydrogen nucleus. The resulting filling factor obtained from the best-fit $n_0,\,v_0$ values and the average mass accretion rate obtained above from our observations is 0.03\%.

\begin{table*}[ht]
\centering
\caption{Integrated fluxes, luminosities and mass accretion rates for the \Hi lines detected in 2M1115 UVES spectrum.}
\renewcommand{\arraystretch}{1.3}
\begin{tabular}{c c c c cc cc}
\hline\hline
Line & order, arm & $F_{\mathrm{line}}/10^{-15}$  & $\log(\Lline)$  & \multicolumn{2}{c}{$\log(\Lacc)$\tablefootmark{b}}  & \multicolumn{2}{c}{$\log(\mdot)$} \\
 &  & (\fluxinteg) & ($L_{\odot}$) & \multicolumn{2}{c}{($L_{\odot}$)} & \multicolumn{2}{c}{(\msun)} \\
\cmidrule(lr){5-6} \cmidrule(lr){7-8}
 &  &   &   & \multicolumn{1}{c}{A21}  & Al17  & \multicolumn{1}{c}{A21} & Al17 \\\hline
H$\alpha$ (H3) & 12, RedU\tablefootmark{a} & $8.43^{+1.61}_{-1.30}$  & $-6.27\pm0.09$ & \multicolumn{1}{c}{$-4.35\pm0.14$ } & $-5.35\pm0.38$ & \multicolumn{1}{c}{$-7.67\pm0.32$ }  & $-8.67\pm0.48$ \\
H$\beta$ (H4) & \phantom{1}2, RedL\tablefootmark{a} & $0.96^{+3.02}_{-0.49}$ & $-7.21\pm0.45$ & \multicolumn{1}{c}{$-4.80\pm0.41$} & $-5.63\pm0.61$ & \multicolumn{1}{c}{$-8.12\pm0.50$} & $-8.95\pm0.67$ \\
H$\gamma$ (H5) & 34-35\tablefootmark{c}, Blue & $0.38^{+2.00}_{-0.28}$ & $-7.61\pm0.46$ & \multicolumn{1}{c}{$-4.87\pm0.42$} & $-5.76\pm0.62$ & \multicolumn{1}{c}{$-8.19\pm0.51$} & $-9.08\pm0.68$ \\
H$\delta$ (H6) & 28, Blue & $1.10^{+3.63}_{-0.53}$ & $-7.15\pm0.45$ & \multicolumn{1}{c}{$-4.24\pm0.41$} & $-5.01\pm0.59$ & $-7.56\pm0.50$ & $-8.33\pm0.66$\\
H$\epsilon$ (H7) & 25, Blue & $0.35^{+4.30}_{-0.26}$ & $-7.65\pm0.48$ & \multicolumn{1}{c}{$-4.44\pm0.43$} & $-5.42\pm0.62$ & \multicolumn{1}{c}{$-7.76\pm0.52$} & $-8.74\pm0.68$ \\\hline
\end{tabular}%
\tablefoot{\\ \tablefoottext{a}{RedU, RedL refers to the upper and lower detectors of the Red arm of UVES respectively.} \\\tablefoottext{b}{\Lacc values were estimated from \Lline using both planetary scaling relations from \citet[][A21]{aoyama2021} for H3--H7, as well as stellar scaling relations from \citet[][Al17]{alcala2017}.} \\ \tablefoottext{c}{The line flux for H$\gamma$ listed here is the average of the fluxes obtained in orders 34 and 35 of the Blue arm.}}
\label{tab2}
\end{table*}

We also estimated accretion rates using \Lline--\Lacc scaling relations from \cite{alcala2017} that were derived empirically based on a sample of young, low-mass stars. Contrary to \cite{aoyama2021}, these relations assume that \Hi emission in case of stellar accretion is coming from gas in the accretion columns that follow the magnetic field lines onto the stellar surface. The \Lacc and \mdot values estimated from the detected \Hi lines based on these stellar scaling relations are also listed in Table~\ref{tab2}. The average mass accretion rate estimated thus for 2M1115 is log(\mdot/\mjyr) $=-8.75\pm0.64$. This is $\sim$1~dex lower than the estimate from \cite{aoyama2021} scaling relations, similar to what was seen in \cite{ringqvist2023} for Delorme~1~(AB)b.

As an independent check, we also derived a mass accretion rate for the target using the 10\% width ($W_{10}$) of its \Ha line. \cite{natta2004} developed an empirical relation between \mdot and \Ha~$W_{10}$, for accretors with $W_{10}\gtrsim200~\kms$ \citep{jayawardhana2003}, using a sample of very low mass objects and T-Tauri stars, spanning a mass range of $0.04 \lesssim \Mstar/\msun \lesssim 0.8$, given by
\begin{equation}
    \log(\mdot) = -12.89(\pm0.3) + 9.7(\pm0.7)\times10^{-3} W_{10}\,,
\end{equation}
where $W_{10}$ is in km\,s$^{-1}$ and \mdot is in \Msyr. Using the measured $W_{10}$ of the overall \Ha profile for the target ($199.4\pm1.2$~\kms), we get an accretion rate of log(\mdot/\mjyr)=$-7.94\pm0.33$, which is consistent with the average \mdot estimate from \cite{aoyama2021} scaling relations within the uncertainties. It is important to note here that values of \mdot and \Ha 10\% width of the sample used in \cite{natta2004} are mostly not simultaneous measurements, and hence the resulting \mdot from the relation should be interpreted with caution. Regardless, the value serves as a cross-check on the mass accretion rate for 2M1115, since it relates the observed \Ha width directly to \mdot instead of relying on a \Lline--\Lacc scaling relation.


\subsection{Likelihood of being bound to 2M1113}
As mentioned in Section \ref{intro}, the M-type star 2MASS J1113+2110 has been suggested to be likely physically associated with 2M1115 based on their spatial proximity and similarity in age and space motion \citep{theissen2018}, with a 1.8\% probability of chance alignment between the two. In light of \textit{Gaia} DR3 observations of these objects, we revisited their properties. Similar to \cite{theissen2018}, neither 2M1115 nor 2M1113 are seen belonging to any of the known stellar associations as implied from the membership probabilities calculated from the Bayesian Analysis for Nearby Young AssociatioNs $\Sigma$ \citep[BANYAN $\Sigma$;][]{gagne2018} tool based on their proper motion, parallax, distance and RV (both show 99.9\% probability to be a field object). With the new \textit{Gaia} distance measurements, the projected physical separation corresponding to the $1.62^{\circ}$ angular separation between the two is 1.28~pc; approximately the same as the distance to Proxima Centauri from Sun \citep[1.29 pc; ][]{turbet2016}. Although the proper motion of 2M1113 \citep[$\mu_{\alpha^*}=-68.932\pm0.194$~mas~yr$^{-1}$, $\mu_{\delta}=-21.985\pm0.190$~mas~yr$^{-1}$][]{gaiadr3} agrees within $\approx1.6$~mas~yr$^{-1}$ of that of 2M1115, their parallax differ by $\approx4.4\sigma$ \citep[parallax $=17.34\pm0.25$~mas for 2M1115;][]{theissen2018}, which reduces the likelihood of the two objects being bound.

To investigate further the possibility that 2M1115 and 2M1113 are physically bound, we calculated the tidal radius of 2M1113 as $R_{\mathrm{tid}}$ = 1.35~pc $\times\,\left(
\Mstar/\msun\right)^{\nicefrac{1}{3}}$ \citep{jiang2010, mamajek2013}. Based on the \cite{theissen2018} mass range for the object (45--75~\mj), we assumed a mass of $60\pm15$~\mj resulting in $R_\mathrm{{tid}}=0.52\pm0.04$~pc. At the distance of 2M1113 ($57.66\pm0.84$~pc) this corresponds to a tidal radius of $0.52\pm0.04^{\circ}$. 2M1115 being at a projected angular separation of $1.62^{\circ}$, is significantly beyond the the tidal influence of 2M1113. Hence the hypothesis that the two are gravitationally bound can be ruled out. However, given the similar proper motion and radial velocities \citep[RV$=-10.1\pm0.3$~\kms for 2M1113;][]{theissen2018} there is a possibility for the two objects to share the same dynamical origin.

\section{Discussion} \label{disc}

From our observations using VLT/UVES at a high resolution $R\sim50,000$ across the optical to near-UV wavelength range, we have detected the first resolved \Hi emission from the young, isolated, L2$\gamma$ dwarf 2MASS J1115+1937, from \Ha to H7. The \Ha emission from the object is broad, with a full width at 10\% peak flux of $199.4\pm1.2$~\kms. This puts 2M1115 right at the 200~\kms threshold for accretion in very low mass stars and brown dwarfs \citep{mohanty2005}. Together with the asymmetries in the observed \Ha, \Hb profiles and the tentative detection of the \Hei emission at 6678~\AA\ (refer Table~\ref{aptab3}), the object displays clear signs of accretion, corroborated by the indicative presence of a disk around the object as seen from the SED (see Figure~\ref{apfig5}). 

\cite{mohanty2005} and \cite{muzerolle2005} presented extensive catalogues of observed \Ha profiles from low-mass stars and sub-stellar objects, along with their mass accretion rates measured from their Ca~\textsc{ii} ($\lambda8662$~\AA) flux in the former, or using radiative transfer modelling based on magnetospheric accretion flows \citep{muzerolle2001} in the latter. Our profile fit for \Ha emission from 2M1115 agrees well with those of the M7-type dwarf CFHT~BD~Tau 4 from \cite{mohanty2005} and Cha~\Ha~1, KPNO~6 of spectral types M7.75, M8.5 respectively, from \cite{muzerolle2005} 
(see Figure~\ref{apfig6}). CFHT~BD~Tau~4, of mass $\sim60~\mj$, is classified as a possible accretor based on its \Ha 10\% width, the presence of a disk and Ca~\textsc{ii} ($\lambda8662$~\AA) emission, with a reported mass accretion rate \mdot$\approx1.3\times10^{-8}$~\mjyr estimated from the latter \citep{mohanty2005}. The model-predicted profile fits in \cite{muzerolle2005} to the accretors Cha~\Ha~1 ($\sim35~\mj$) and KPNO~6 ($\sim25~\mj$) indicate mass accretion rates of $5.25\times10^{-9}$ and $4.17\times10^{-9}$~\mjyr respectively. The average mass accretion rate obtained for 2M1115 using \cite{aoyama2021} scaling relations ($1.4^{+2.8}_{-0.9}\times10^{-8}$~\mjyr) agree within $\sim1.1\sigma$ with the estimated \mdot for all the three objects, showing consistency with the matching line profiles. The \mdot obtained based on the \Ha~\Wten \citep{natta2004}, $1.1^{+1.3}_{-0.6}\times10^{-8}$~\mjyr, also agrees well with those of these three accretors. On the other hand, the average \mdot obtained from the \citep{alcala2017} scaling relations ($1.8^{+6.0}_{-1.4}\times10^{-9}$~\mjyr) is an order of magnitude lower than that of CFHT BD Tau 4, but agrees with those of Cha \Ha~1 and KPNO 6.

The associated accretion luminosity derived for 2M1115 from \cite{aoyama2021} scaling relations is, on average, $\mathrm{log}(L_{\mathrm{acc}}/L_{\odot})=-4.54\pm0.38$. The \Teff--dependent chromospheric noise limit from \cite{manara2013} \citep[see also][]{venuti2019} derived from a sample of non-accreting young stellar objects with \Teff$\leq4000$~K is given by
\begin{equation}
    \mathrm{log}(L_{\mathrm{acc},\,\mathrm{noise}}/\Lbol)=6.2(\pm0.5)\times\mathrm{log}(T_{\mathrm{eff}}) - 24.5(\pm1.9).
\end{equation}
This limit describes the simulated luminosity from a non-accreting young star of a certain spectral type (or \Teff) solely due to its chromospheric activity.
The $\Teff=1816\pm63$~K we derived translates into a chromospheric noise limit of $\mathrm{log}(L_{\mathrm{acc,\,noise}}/L_{\mathrm{bol}})=-4.29\pm2.50$ for 2M1115. Together with its estimated \Lbol from SED fit, this corresponds to a luminosity limit of $\log(L_{\mathrm{acc},\,\mathrm{noise}}/\lsun)=-7.92\pm2.50$. The estimated accretion luminosity from its \Hi emission is higher by more than three orders of magnitude (3.4~dex) compared to this noise limit. Although the relation is derived from K and M-type stars and may have uncertainties when extrapolated to planetary temperatures, it is unlikely that the true $L_{\mathrm{acc,\,noise}}$ at these temperatures are higher by several orders of magnitude than what is derived from the extrapolation, to reach the \Lacc level estimated for 2M1115 from these observations. Thus, there is very less likelihood that all the \Hi lines from 2M1115 are solely due to its chromospheric activity. 

Among the detected \Hi lines, H6 and H7 show similar velocity shifts (within uncertainties) as the systemic RV of $-14\pm7$~\kms (refer Table~\ref{aptab2}), along with H5 in order 35 and \Hb in order 3. H5 in order 34 is slightly blue-shifted relative to the systemic velocity. The \Hb profile in order 2 has both broad and narrow components, which are blue-shifted ($\sim-14$~\kms) and red-shifted ($\sim+8$~\kms) relative to the systemic RV. The \Ha profile has two NC in addition to the BC. The BC is red-shifted relative to the system velocity, as predicted by the \cite{aoyama2018, aoyama2021} models (discussed in Section~\ref{assymetry}), along with a blue-shifted NC; however the red-shifted NC cannot be explained within the planetary accretion shock framework. The \Ha emission is also significantly stronger than other detected \Hi lines, and does not agree with predicted flux from planet-surface shock model that fits the rest of the \Hi lines very well. This could, as discussed in Section~\ref{assymetry}, point to a partial or even dominating contribution from mechanisms other than shock-heating for \Ha emission like heating along the accretion columns or chromospheric activity. The best-fit planet-surface shock model that fits all detected \Hi lines other than \Ha predicts a filling factor of 0.03\%, which hints at a magnetospheric accretion geometry for 2M1115 \citep{ringqvist2023}. This provides a tentative indirect evidence for a magnetic field on such a low-mass and young isolated object.

Moderate-resolution SDSS spectra available for the target from DR9 (March 2007) and DR12 (April 2012) epochs, together with the UVES epoch (June 2023) data, facilitate an analysis of the accretion over time. From Tables~\ref{tab2} and \ref{aptab5}, it can be inferred that the H$\alpha$ flux decreases by $\sim7$\% between 2007 and 2012, and then increases by $\sim20$\% between 2012 and 2023. On the other hand, line fluxes for \Hb--H7 increases by $200-300$\% between 2007 and 2012, and decreases by $\sim50-80$\% between 2012 and 2023 for \Hb, H5 and H7, and by $\sim15\%$ for H6. 
On average, the \Hi accretion luminosity in the DR12 epoch ($\log(\Lacc/\lsun)=-4.30\pm0.14$) is higher by $\sim150$\% from the DR9 epoch ($\log(\Lacc/\lsun)=-4.68\pm0.14$) and by $\sim75$\% from the UVES epoch ($\log(\Lacc/\lsun)=-4.54\pm0.38$). The average mass accretion rates estimated from all the \Hi lines in all three epochs imply the same progression, varying as \mdot$\approx1\times10^{-8}$, $2.4\times10^{-8}$ and $1.4\times10^{-8}$~\mjyr between the DR9, DR12 and UVES epochs respectively. The trend seen from the hydrogen emission over the three epochs suggests a variability in accretion in the target, and could benefit from a future detailed time-resolved monitoring of the emission lines.

A major source of uncertainty in the analysis presented in this work is the loosely constrained mass of the target. The previously reported mass for the target in the literature  \citep[7--21~\mj;][]{theissen2018} was estimated from evolutionary models using the temperature estimate from SED fit to its \textit{zJHK$_S$W1} photometry based on the photometric distance estimate of 37~pc, and the age range 5--45~Myr. The lower age limit was roughly based on the end of the accretion phase for brown dwarfs \citep{mohanty2005} and the upper age limit was taken as that of the oldest accreting object known to harbour primordial circumstellar material \citep{boucher2016}. With the revised distance and resulting SED fit discussed in Section~\ref{mar}, we attempted to constrain the mass further. Using the updated absolute magnitudes in the $J$, $H$, $K_S$ bands, from the DUSTY00 atmosphere isochronal models for very low mass stars and brown dwarfs \citep{chabrier2000} we infer a mass of $\sim7.5$~\mj at 5~Myr and $\sim21$~\mj at 50~Myr. Alternatively, we also interpolated the AMES-Cond \citep{allard2001} and BT-Settl \citep{allard2012, allard2013} model isochronal grids for the temperature from our SED fit (1816~K) in the age range 5--50~Myr. This gives a slightly lower and stricter mass range of 9--15~\mj. 
The surface gravity ($\log(g)=3.83\pm0.24$) and radius from the best-fit atmospheric models to the SED (refer Appendix~\ref{sed}) imply a slightly lower mass range of $6\pm4$~\mj, while the best-fit \cite{aoyama2018,aoyama2021} model to the observed \Hi predicts a mass $\Mp=6^{+8}_{-4}$~\mj (refer Section~\ref{mar}). The latter estimate is not quite independent from the former since it is calculated partly based on the radius estimate from the SED fit. Nevertheless, the implied mass range from our analysis is slightly lower than the existing mass range in the literature, but still remains fairly large. Depending on the lowest (2~\mj) to highest (21~\mj) value in the mass range listed for the target here, the estimated mass accretion rate (\mdot/\mjyr) changes by a factor of approximately 1 dex. 

Our observations of 2M1115 and the resulting findings demonstrate the great potential of VLT/UVES for studying young, accreting planets. We find that 2M1115 could greatly benefit from further study; better constraints on its age could be placed by obtaining photometry covering a longer wavelength range in its SED. A dynamic, model-independent estimate of its mass can be obtained by studying the disk kinematics using ALMA \citep{alma}. Imaging in the mid-infrared using instruments like JWST/MIRI \citep{miri} to study the spectral distribution of 2M1115 at the mid-infrared wavelengths could also help in constraining the properties of the disk. Further, spectroscopic observations of 2M1115 in the near-infrared to study accretion signatures such as Paschen and Brackett emission lines can serve as an independent check on the mass accretion rate and give further insights into the accretion geometry.
Monitoring emission lines from the target over various timescales with UVES or other upcoming high-resolution spectrographs like ANDES \citep[formerly known as HIRES;][]{hires} can help investigate accretion variability. 
Finally, continued monitoring of 2M1115's space motion could help solve the mystery of whether it belongs to yet undiscovered stellar associations or has been ejected from currently known ones.

\begin{acknowledgements}
M.J.\ gratefully acknowledges funding from the Knut and Alice Wallenberg Foundation and the Swedish Research Council.
G.-D.M.\ acknowledges the support of the DFG priority program SPP 1992 ``Exploring the Diversity of Extrasolar Planets'' (MA~9185/1) and from the Swiss National Science Foundation under grant 200021\_204847 ``PlanetsInTime''.
Parts of this work have been carried out within the framework of the NCCR PlanetS supported by the Swiss National Science Foundation.
This publication makes use of VOSA, developed under the Spanish Virtual Observatory (\url{https://svo.cab.inta-csic.es}) project funded by MCIN/AEI/10.13039/501100011033/ through grant PID2020-112949GB-I00. 
VOSA has been partially updated by using funding from the European Union's Horizon 2020 Research and Innovation Programme, under Grant Agreement nº 776403 (EXOPLANETS-A).
Funding for SDSS-III has been provided by the Alfred P. Sloan Foundation, the Participating Institutions, the National Science Foundation, and the U.S. Department of Energy Office of Science. The SDSS-III web site is \url{http://www.sdss3.org}. SDSS-III is managed by the Astrophysical Research Consortium for the Participating Institutions of the SDSS-III Collaboration including the University of Arizona, the Brazilian Participation Group, Brookhaven National Laboratory, Carnegie Mellon University, University of Florida, the French Participation Group, the German Participation Group, Harvard University, the Instituto de Astrofisica de Canarias, the Michigan State/Notre Dame/JINA Participation Group, Johns Hopkins University, Lawrence Berkeley National Laboratory, Max Planck Institute for Astrophysics, Max Planck Institute for Extraterrestrial Physics, New Mexico State University, New York University, Ohio State University, Pennsylvania State University, University of Portsmouth, Princeton University, the Spanish Participation Group, University of Tokyo, University of Utah, Vanderbilt University, University of Virginia, University of Washington, and Yale University.
\end{acknowledgements}

\bibliographystyle{aa.bst} 
\bibliography{bibliography}
 
\begin{appendix}
\section{Details of observations}

Table~\ref{aptab1} gives the observing log for 2M1115 with VLT/UVES during 2023 June 10--11, as part of the ESO programme 0111.C-0166(A).  The integration time was 740~s for each observation with the number of integrations, NDIT=1. The seeing measurements quoted here are estimated from the FWHM of the 2D spatial PSF of the \Ha line from the data.

\begin{table}[!htb]
\centering
\caption{Observing log for the VLT/UVES observations of 2M1115 on 2023 June 10, 11.} 
\begin{tabular}{c c c c}
\hline\hline
Date & UT Time & Seeing & Air mass \\
(MJD) & (hh:mm) & ($\arcsec$) & \\
\hline 
   60105.96795718 & 23:13 & 1.504 & 1.408 \\
   60105.97697917 & 23:26 & 1.350 & 1.424 \\
   60106.96561343 & 23:10 & 1.401 & 1.408 \\
   60106.97460648 & 23:23 & 1.451 & 1.425 \\
\hline
\end{tabular}
\label{aptab1} 
\end{table}

\section{Flux calibration}\label{apendixb}
The low resolution spectrum from SDSS DR12 \citep{sdss12} and DR9 \citep{sdss9} does not reveal any significant continuum for 2M1115 in the relevant wavelength range for the UVES data. The stacked UVES spectrum when binned to $\sim5$~\AA\ resolution also does not show any continuum, except for a faint continuum that begins to appear near the higher wavelength end of the upper red arm ($>6000$~\AA). With no considerable continuum emission from the object, flux calibration based on a theoretical model fit to the object's SED would not be ideal since the photometric bands in the optical wavelength range will be mostly dominated by the flux from the emission lines. 

An alternative method would be to calibrate the object's flux based on the observing instrument's response curve. For this purpose, we use the observations of the standard star LTT 6248 taken with UVES on 2023 June 11 with the same observing set-up as 2M1115. A well-calibrated spectrum for this star exists in the literature \citep{rubin2022} in the required wavelength range. The UVES spectrum for LTT 6248 was extracted using the default ESO pipeline and was corrected for atmospheric transmission at an air mass of 1.7 corresponding to its observing conditions, using the sky transmission curve obtained from ESO's SKYCALC Sky Model Calculator\footnote{\url{https://www.eso.org/observing/etc/bin/gen/form?INS.MODE=swspectr+INS.NAME=SKYCALC}}. The instrument response for UVES was derived by dividing the spectrophotometric calibrated spectrum of LTT 6248 from \cite{rubin2022} with its uncalibrated UVES spectrum normalised for the corresponding integration time (500 s). 

We then flux calibrated the 2M1115 UVES spectrum by first correcting it for atmospheric transmission at the average air mass of its observations (1.415), then normalising it for its integration time of 740 s and finally multiplying it by the instrument response curve derived above. The resulting calibrated spectrum for the target shows consistency between the integrated flux for \Hi emission lines with those from earlier SDSS epochs (refer Table~\ref{aptab4}). 


\section{Identification of potential lines}\label{apendixc}
Potential emission lines were selected from the stacked spectrum and classified as confirmed or tentative detection based on a detection confidence determined as described here. For each potential line identified in the spectrum, a region of 20~\AA\ width is defined around the line, usually centred on the line except when the line is near the beginning/end of an order (in which case the region is shifted by 5~\AA\ away from the noisier part). The local noise is calculated as the standard deviation within this region, avoiding $\pm2$~\AA\ around the centre of the line. A single Gaussian is then fit to the line and the detection confidence is determined as the strength of the line, estimated as the peak of the resulting Gaussian fit. Identified emission lines are classified as confirmed detections if the corresponding detection confidence is higher than $5\sigma$, with $\sigma$ being the local noise. We note that a high threshold of $5\sigma$ is used since we do fit for a continuum in the spectrum (as discussed in Appendix~\ref{apendixb}) and thus may be underestimating the corresponding noise. Alternatively, if the line is recurrent in two subsequent orders, and the rms of the two line strengths is higher than $5\sigma$, it is also classified as a confirmed detection. Lines identified similarly with a confidence of 3--$5\sigma$ are classified as tentative detections.

\section{Characteristics of \Hi line profiles} \label{HIchar}
\begin{figure*}[!htb]
    \centering
    \includegraphics[width=0.49\linewidth]{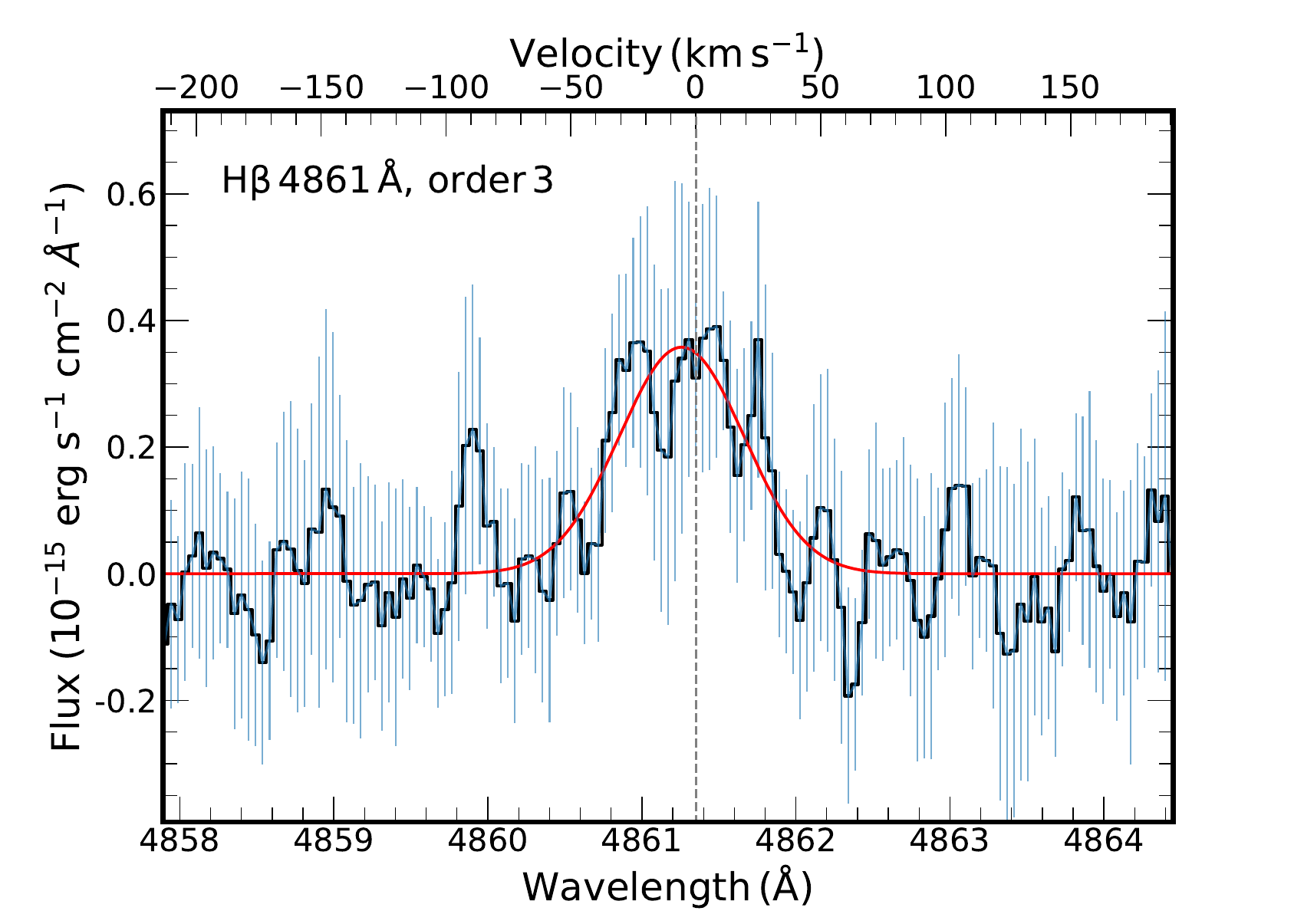}
    \includegraphics[width=0.49\linewidth]{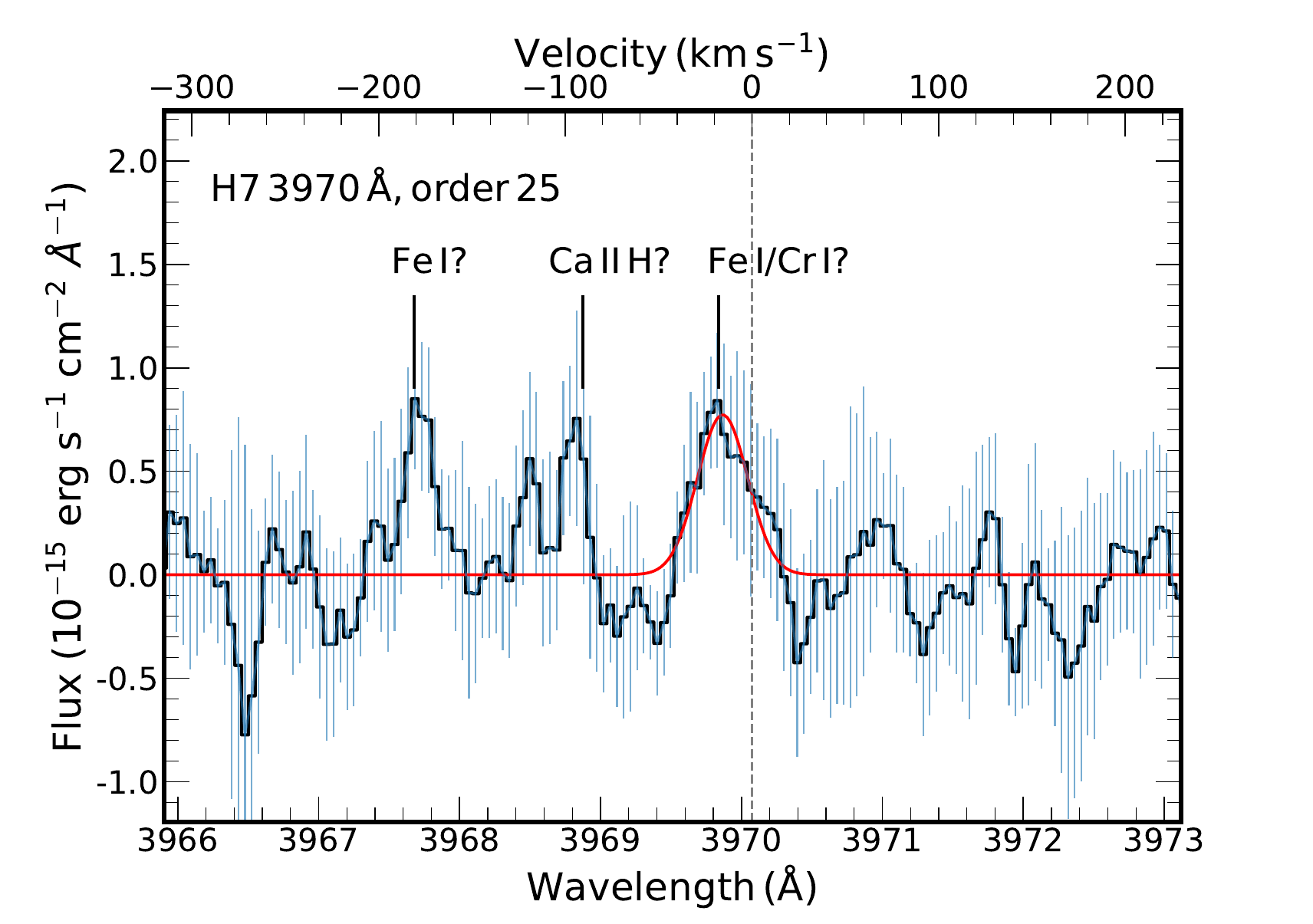}
    \caption{(Left) H$\beta$ detected at $4.3\sigma$ in order 3 of the RedL arm, and (Right) H7 line detected at $3.6\sigma$ in order 25 of the Blue arm of UVES in the 2M1115 data. The red line indicates the Gaussian fit to the observed flux (blue line). The vertical lines indicate the uncertainty in the flux. The tentatively detected Ca~\textsc{ii}~H (3968.47~\AA) at $3.1\sigma$ and Fe~\textsc{i} (3967.42~\AA) at $3.2\sigma$ next to H7 are also indicated in the figure.}
    \label{H3-H7}
\end{figure*}

\begin{figure*}[!ht]
    \centering
    \includegraphics[width=0.49\linewidth]{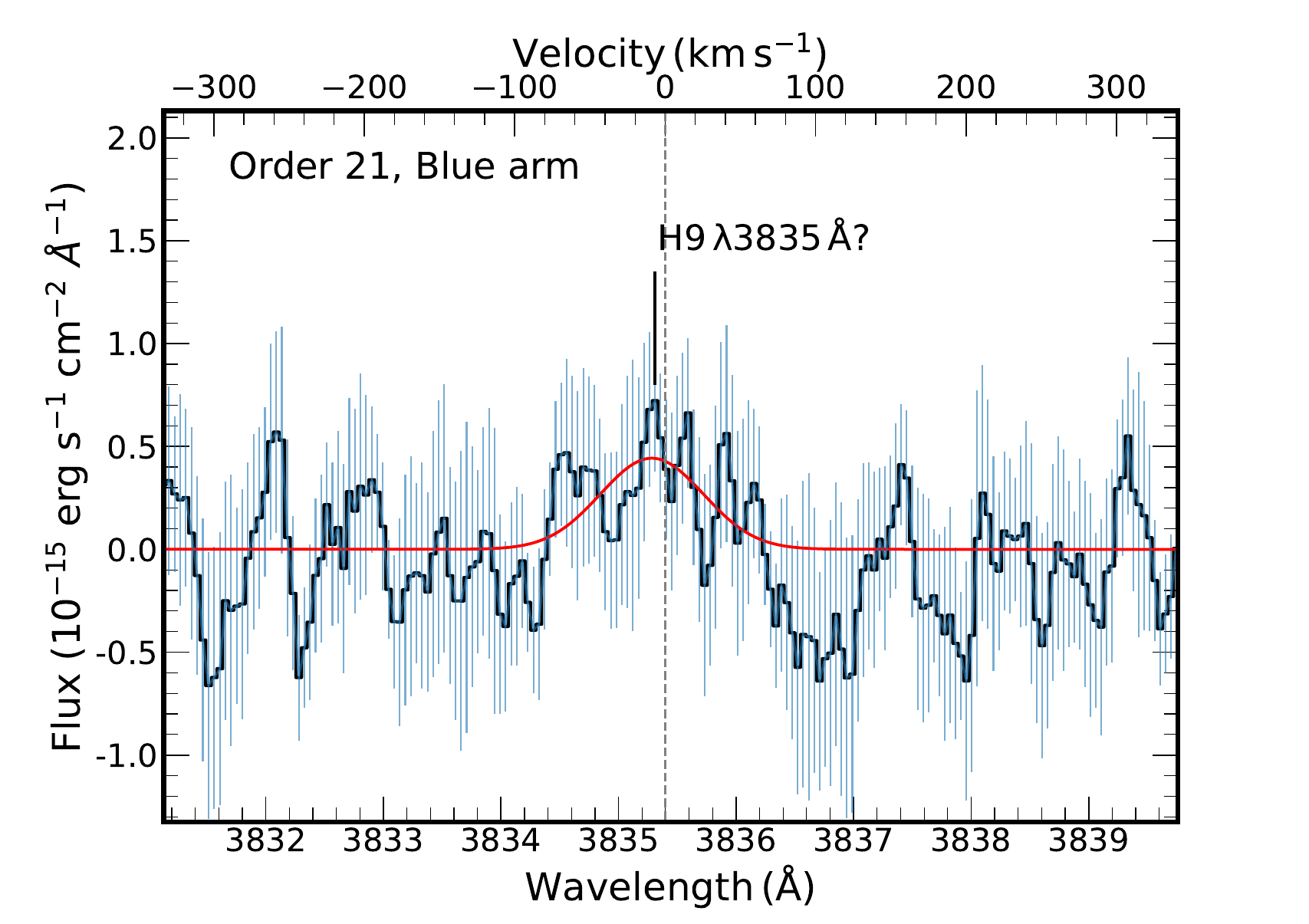}
   \includegraphics[width=0.49\linewidth]{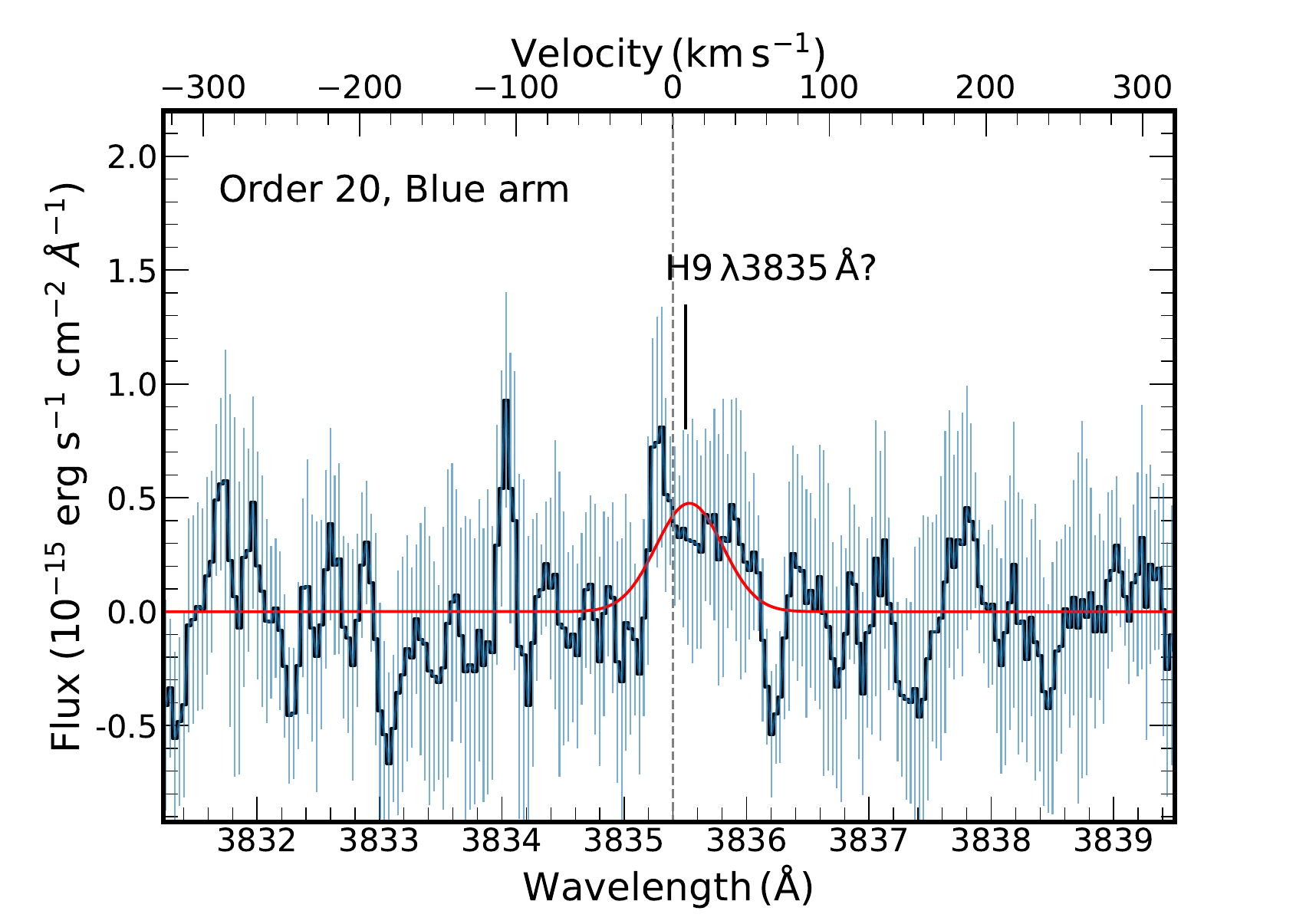}
    \caption{The flux increase detected at the location of H9 in (Left) order 21, and (Right) order 20 of the Blue arm of UVES in the 2M1115 data, at $1.7\sigma$ and $1.9\sigma$ respectively. Colours and symbols hold the same meaning as in the previous figure.}
    \label{H9}
\end{figure*}

Figure~\ref{H3-H7}
illustrate the Gaussian fits to the profile of the tentatively detected H$\beta$ in order 3 of the RedL arm ($4.3\sigma$) and H7 in order 25 of the Blue arm ($3.6\sigma$). H7 shows a slight increase in flux at the blue side of the line, indicative of a possible blending with either Fe~\textsc{i} at 3970.63~\AA\ or Cr~\textsc{i} at 3969.74~\AA. H7 was also detected in order 24, but since the detection confidence is only $1.7\sigma$, we do not include it in our analysis. We also detected a clear flux increase in orders 20 and 21 of the Blue arm corresponding to the H9 line at 3835.40~\AA\ (see Figure~\ref{H9}) but the detection strength for both were not formally significant ($1.9\sigma$ and $1.7\sigma$ respectively) and has not been listed among detections in this work.
The characteristics of the line profiles for all the \Hi emission lines detected in the spectrum are given in Table~\ref{aptab2}. \Ha and \Hb are best explained with a triple (t) and double (d) Gaussian fits respectively while the rest of the \Hi lines, being of lower S/N, are fitted with single (s) Gaussians.

\begin{table*}
\tiny
\centering
\caption{Characteristics of the line profiles for \Hi emission lines detected in the 2M1115 UVES spectrum.}
\resizebox{0.9\textwidth}{!}{
\begin{tabular}[l]{c c c c c c c c}
\hline\hline
Line & \lamrest  & Det. conf.\tablefootmark{a} & Fit\tablefootmark{b} & \lamobs & RV & FWHM & \Wten\\
 & (\AA) & ($\sigma$) &  & (\AA) & (\kms) & (\AA) & (\AA)\\\hline
\Ha (H3) & $6562.79$ & 59.1 & t, BC & $6562.68\pm0.02$ &  $-5.04\pm2.48$ & $2.97\pm0.08$ & $5.38\pm0.09$\\ 
& & & t, NC$_1$ & $6562.21\pm0.05$ &  $-26.24\pm3.20$ & $0.99\pm0.03$ & $1.76\pm0.01$ \\
& & & t, NC$_2$ & $6563.15\pm0.03$ & $16.34\pm2.66$ & $0.92\pm0.03$ & $1.64\pm0.03$\\
H$\beta_2$ (H4$_2$) & $4861.35$ & 8.7 & d, BC & $4860.90\pm0.55$ &  $-28.02\pm33.71$ & $1.36\pm0.24$ & $2.45\pm0.45$\\
& & & d, NC & $4861.26\pm0.17$ &  $-5.50\pm10.79$ & $0.80\pm0.10$ & $1.43\pm0.16$ \\
H$\beta_3$ (H4$_3$) & $4861.35$ & 4.3 & s & $4861.26\pm0.03$ & $-5.67\pm2.66$ & $0.95\pm0.07$ & $1.73\pm0.03$\\
H$\gamma_{34}$ (H5$_{34}$) & $4340.47$ & 4.6 & s & $4340.44\pm0.02$ & $-2.14\pm1.37$ & $0.45\pm0.04$ & $0.80\pm0.01$\\
H$\gamma_{35}$ (H5$_{35}$) & $4340.47$ & 4.6 & s & $4340.39\pm0.04$ & $-5.57\pm2.52$ & $0.95\pm0.08$ & $1.73\pm0.03$\\
H$\delta$ (H6) & $4101.73$ & 7.0 & s & $4101.64\pm0.02$ & $-6.86\pm1.61$ & $0.88\pm0.05$ & $1.60\pm0.02$ \\
H$\epsilon$ (H7)\tablefootmark{c} & $3970.07$ & 3.5 & s & $3969.87\pm0.03$ & $-15.49\pm2.55$ & $0.42\pm0.08$ & $0.75\pm0.02$ \\\hline
\end{tabular}
}
\tablefoot{\\ \tablefoottext{a}{`Det.\ conf.' stands for detection confidence, which is expressed in units of $\sigma$, the standard deviation within $\pm20$~\AA\ around the line (refer Section~\ref{apendixc}).}\\ \tablefoottext{b}{(s), (d) and (t) denote single, double and triple Gaussian fits respectively, and BC and NC denote broad and narrow Gaussian components.}\\ \tablefoottext{c}{H7 is tentatively detected.}}
\label{aptab2}
\end{table*}

\section{Analysis with planet-surface shock model} \label{shockmodel}
We fit the observed H3--H7 line profiles of 2M1115 with the \cite{aoyama2020, aoyama2021} model grids assuming \Hi line emission from shock at the planet surface.
\begin{figure}[!hb]
    \centering
    \includegraphics[width=1\linewidth]{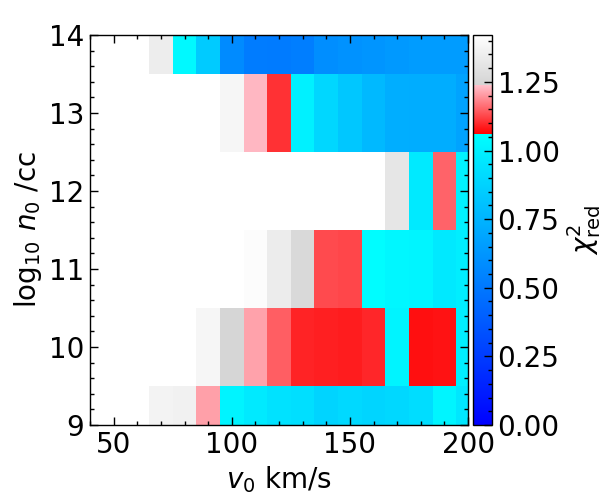}
    \caption{ Results from fitting \cite{aoyama2020, aoyama2021} model-grid to the observed H4--H7 line profiles of 2M1115. The model-grid allows $n_0$ to range from $10^9-10^{14}$~cm$^{-3}$ and $v_0$ to range from $50-200~\kms$. The corresponding $\chi_\mathrm{red}^2$ is represented by the colour bar, with $\chi_\mathrm{red}^2=1$ indicating a good fit.}
    \label{apfig3}
\end{figure}

\begin{figure*}[!ht]
    \centering
    \includegraphics[width=0.9\linewidth]{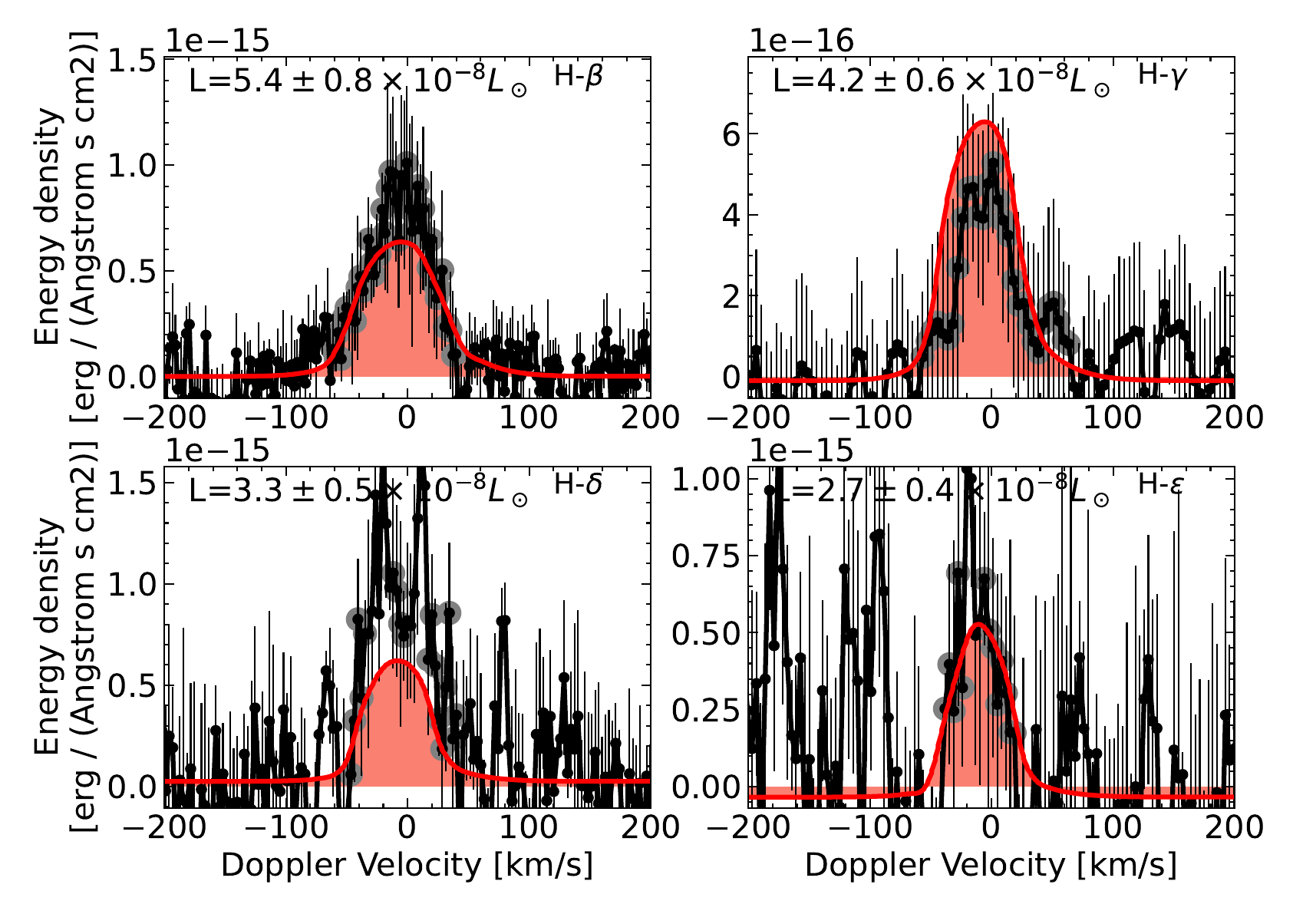}
    \caption{The SED with \cite{aoyama2020, aoyama2021} model corresponding to $\chi_\mathrm{red}^2\leq1$ that fits all the line profiles (H4--H7) reasonably well. The y-axis represents the measured flux density and the x-axis represents the velocity shift with respect to the rest wavelength of the line. The solid black curve represents the observed data, and the red curve represents the model prediction. The black vertical lines are the uncertainties in the observed flux, and the grey filled circles are the points along the observed profile the model was fitted to. The model-predicted luminosity from each line is also indicated in the figure.}
    \label{bestshockmodel}
\end{figure*}

 The quality of the fit was determined by reduced $\chi^2$ ($\chi_\mathrm{red}^2$) with $\chi^2$ defined as 
\begin{equation}
    \chi^2 = \sum_i \left(\frac{F_\mathrm{model,\,i} - F_\mathrm{obs,\,i}}{\sigma_i} \right)^2.
\end{equation}
Here, $\sigma$ is the uncertainty in the observed flux $F_\mathrm{obs,\,i}$ for bin $i$ and $F_\mathrm{model,\,i}$ is the corresponding flux prediction from the model. The value of $\chi_\mathrm{red}^2$ for the fit is then obtained by dividing $\chi^2$ by the degree of freedom, which is the number of bins used in the fitting process. A $\chi_\mathrm{red}^2\approx1$ (light blue) indicates good agreement between the model prediction and the observed profile, and $\chi_\mathrm{red}^2 < 1$ (darker blue) indicates that the difference is less than the uncertainty in the flux. 

\Ha is significantly brighter than the rest of the \Hi lines detected; trying to fit for \Ha using a high extinction of 3.5~mag shows very poor agreement with H5, H6 and H7 profiles. Hence, \Ha has been excluded from the following analysis. For fitting H5, we took the average of its line profiles from orders 34 and 35. For H5--H7, we also masked the bins corresponding to possible contamination from metal lines, as described in Section~\ref{profilechar} and Appendix~\ref{HIchar}. The resulting quality of the fit improved greatly, with good agreement over a wider parameter range (see Figure~\ref{apfig3}) as compared to when these possible metal lines were included in the fitting process. This highlights the importance of the high spectral resolution such as with UVES, which helps identify overlapping metal lines in the line profile. 

Among the models that show good agreement with $\chi_\mathrm{red}^2\leq1$, adopting a pre-shock gas velocity of $v_0=120$~\kms and a pre-shock gas density of $n_0=10^{14}$~\cmcube reproduces all the line profiles (H4--H7) very well (see Figure~\ref{bestshockmodel}), while others reproduce some of the lines well but not others.  The blue region in the figure indicate those values of $v_0,\,n_0$ for which fitting residuals are not statistically significant at the $1\sigma$ level. The pre-shock gas parameters over this region range from $v_0=80-200~\kms$ and $n_0=10^9-10^{14}~\cmcube$. Accordingly, we define the uncertainties in the best-fit model parameters as $v_0=120^{+80}_{-40}$~\kms and $\log (n_0/\textrm{cm}^3) = 14_{-5}^{+0}$. The upper uncertainty in  $n_0$ is set as 0 since the best-fit value reaches the edge of the model grid. 

\section{Comparison of H3 with standard line profiles in the literature} \label{H3comp}
\begin{figure}
    \centering
    \includegraphics[width=1.0\linewidth, trim = {0.5cm 0cm 0.2cm 0.5cm}, clip]{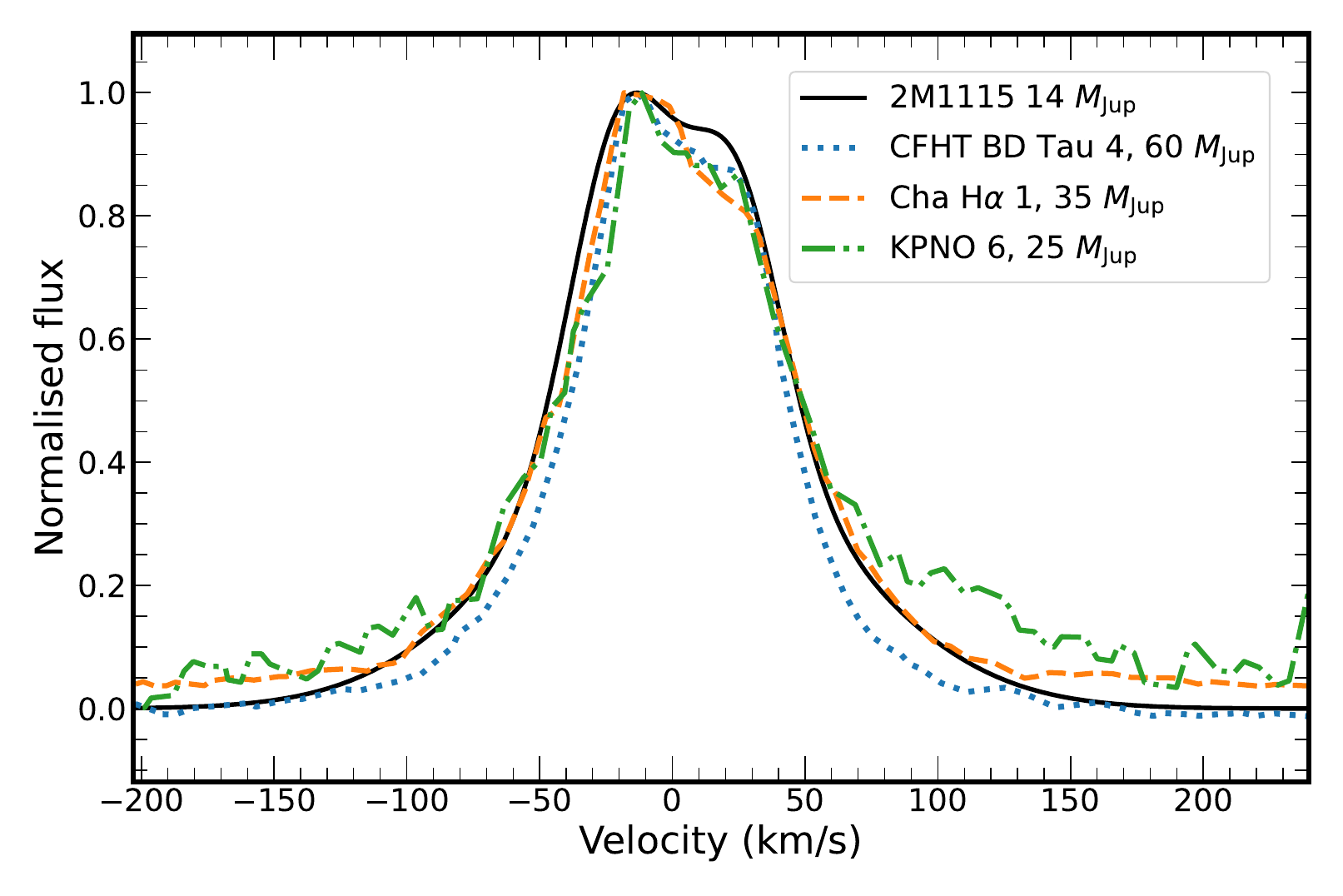}
    \caption{Profile fit to \Ha emission from 2M1115 compared to normalised \Ha profiles for CFHT~BD~Tau~4 from \cite{mohanty2005} and Cha~\Ha~1, KPNO~6 from \cite{muzerolle2005}. The \Ha profile for 2M1115 in this plot has been corrected for its observed velocity shift.}
    \label{apfig6}
\end{figure}

The comparison of the obtained \Ha line profile for the target with the best-matching \Ha profiles from \cite{muzerolle2005} and \cite{mohanty2005} is illustrated in Figure~\ref{apfig6}. The line profile for all the objects have been normalised by their peak flux for the sake of comparison. The \Ha profile for 2M1115 in this plot has been corrected for the line's observed radial velocity shift ($-7.3\pm1.4$~\kms) as determined from a single Gaussian fit to its line profile.

\section{Non-Hydrogen line detections}
\begin{figure*}
\centering
\includegraphics[width=0.49\linewidth]{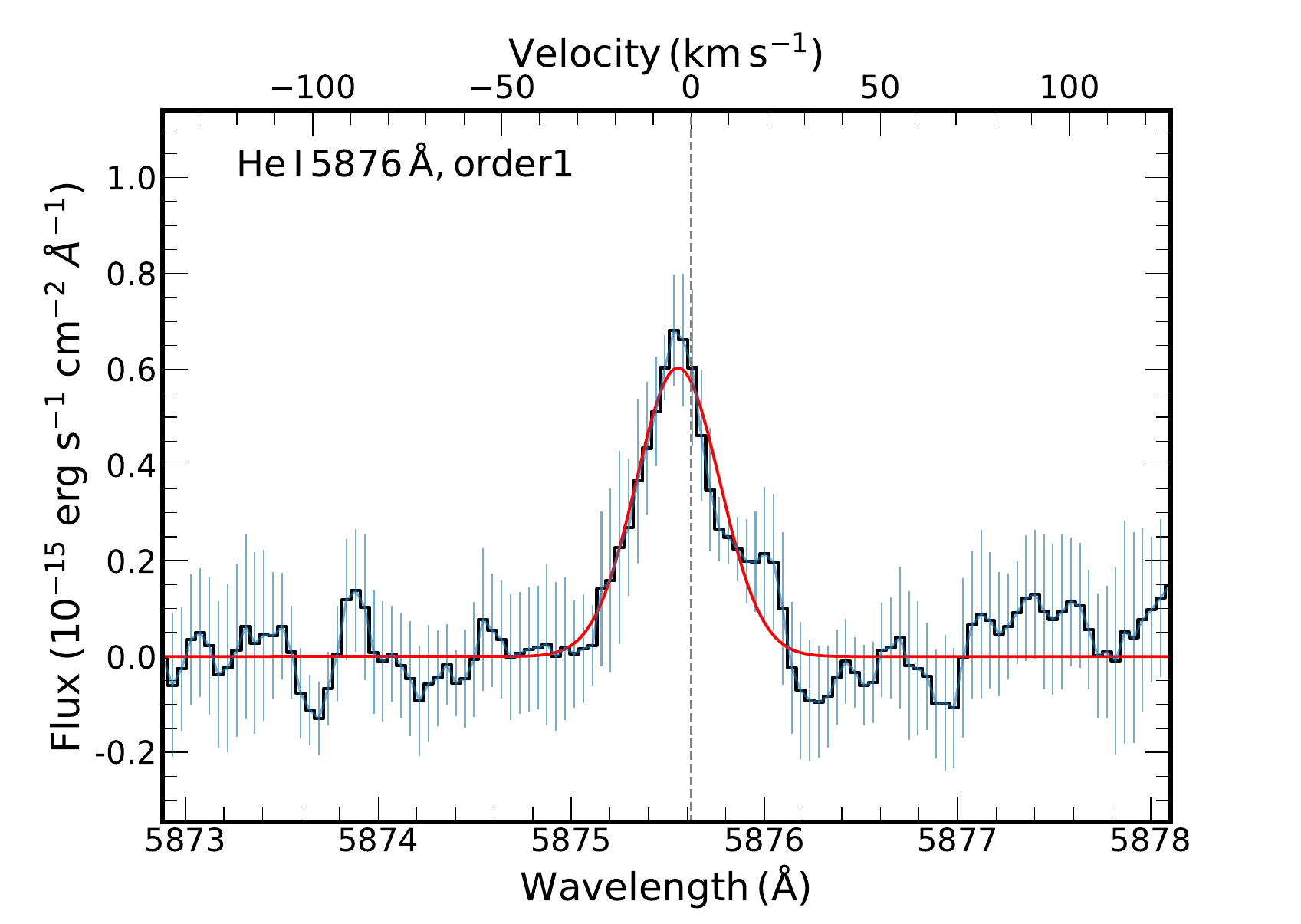}
\includegraphics[width=0.49\linewidth]{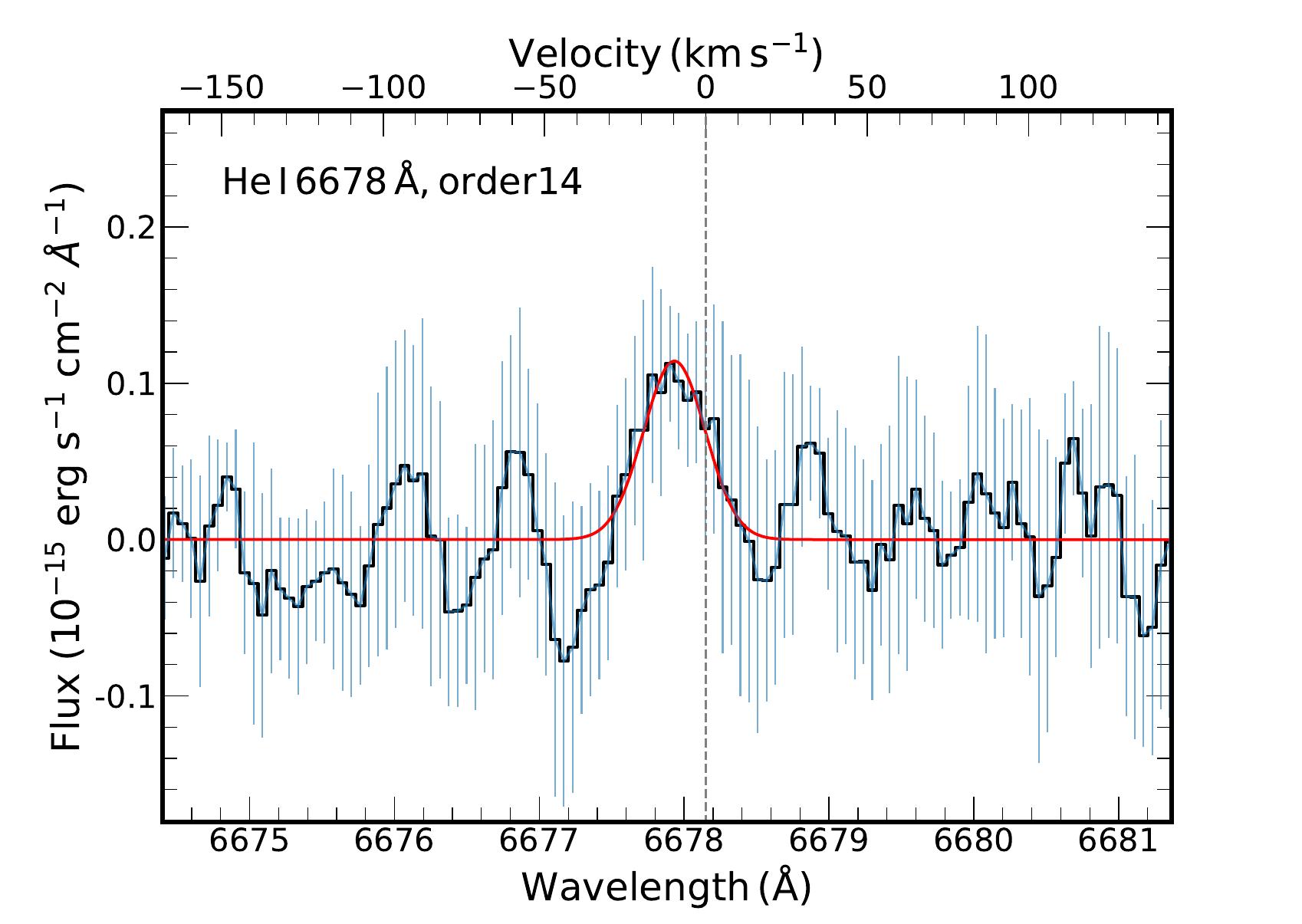}
\caption{Line profiles of \Hei at (Left) 5876~\AA\ and (Right) 6678~\AA\ detected in the order 1 and 14 of the UVES RedU arm at $8.6\sigma$ and $3.2\sigma$ respectively in the 2M1115 data. Colours and symbols hold the same meaning as in previous figures.}
\label{HeI}
\end{figure*}

Apart from \Hi lines, we also detected \Hei emission (see Figure~\ref{HeI}) and several metal lines including Ca~\textsc{ii}~H, Fe~\textsc{i}, Cr~\textsc{i} and Ti~\textsc{i}. \Hei emission at 5876~\AA\ was detected at high confidence ($8.6\sigma$) while \Hei emission at 6678~\AA\ was detected tentatively ($3.2\sigma$).
Ca~\textsc{ii}~H was detected tentatively in order 25 of the Blue arm (see Figure~\ref{H3-H7}, right panel) at $3.4\sigma$; the line also recurred in order 24 but at only $1.6\sigma$ with no formal significance to include in the detection list. Both the Ca~\textsc{ii}~H detections were at an RV of $\sim25$~\kms, much redder compared to the systemic RV of $-14\pm7\,\kms$.
Among the possible metal lines detected (all $<5\sigma$), Fe~\textsc{i} emission lines at $\lambda3967.42$, 4970.50, 4986.22, and 5684.43~\AA\ were detected at similarly redder RV shifts ($\sim17$--$35\,\kms$), while Fe~\textsc{i} emission lines at 4839.54 and 6703.57~\AA\ were detected at bluer RV shifts ($\sim-25$~\kms) compared to the systemic velocity. We note that this is different from the expected behaviour of metal lines as seen from the stellar cases, which adhere more or less to the systemic velocity \citep{sicilia2015}.
Table~\ref{aptab3} lists the characteristics of the line profiles for all the non-hydrogen lines detected in the data.

\begin{table*}[ht]
\centering
\caption{Characteristics of the line profiles for confirmed and tentative non-Hydrogen lines detected in the 2M1115 UVES spectrum.}
\begin{tabular}[l]{c c c c c c c c}
\hline\hline
Line\tablefootmark{a} & \lamrest & order, arm & Det. conf. & \lamobs & RV & FWHM & \Wten \\
 & (\AA) & & ($\sigma$) & (\AA) & (\kms) & (\AA) &(\AA) \\\hline
 & & \multicolumn{4}{c}{Confirmed Detections} & & \\
\hline
\Hei & $5875.62$ & \phantom{1}1, RedU & 8.6 & $5875.55\pm0.01$ & $\,\,-3.53\pm0.61$ & $0.51\pm0.03$ & $0.91\pm0.01$ \\
\hline
& & \multicolumn{4}{c}{Tentative Detections} & & \\
\hline
Fe~\textsc{i} & $3834.22$ & 20, Blue & 3.5 & $3834.03\pm0.01$ & $-14.62\pm1.02$ & $0.10\pm0.03$ & $0.18\pm0.01$ \\
Fe~\textsc{i} & $3967.42$ & 25, Blue & 3.2 & $3967.72\pm0.03$ & $\phantom{1}22.47\pm2.15$ & $0.26\pm0.07$ & $0.47\pm0.02$ \\
Ca~\textsc{ii}~H$_{25}$\tablefootmark{b} & $3968.47$ & 25, Blue & 3.4 & $3968.81\pm0.02$ & $\phantom{1}25.58\pm1.86$ & $0.17\pm0.06$ & $0.30\pm0.02$ \\ 
Fe~\textsc{i} & $4839.54$ & \phantom{1}2, RedL & 3.1 & $4839.13\pm0.01$ & $-25.47\pm0.96$ & $0.12\pm0.04$ & $0.21\pm0.02$ \\
Fe~\textsc{i} & $4970.50$ & \phantom{1}5, RedL & 3.5 & $4970.78\pm0.01$ & $\phantom{1}16.93\pm0.85$ & $0.12\pm0.03$ & $0.21\pm0.01$ \\
Fe~\textsc{i} & $4986.22$ & \phantom{1}6, RedL & 4.8 & $4986.80\pm0.01$ & $\phantom{1}34.66\pm0.78$ & $0.14\pm0.03$ & $0.26\pm0.01$ \\
Fe~\textsc{i} & $5329.99$ & 14, RedL & 3.4 & $5329.71\pm0.02$ & $-15.55\pm1.08$ & $0.19\pm0.04$ & $0.34\pm0.01$ \\
Fe~\textsc{i} & $5684.43$ & 21, RedL & 4.4 & $5684.81\pm0.02$ & $\phantom{1}17.44\pm0.91$ & $0.22\pm0.04$ & $0.40\pm0.01$ \\
Ti~\textsc{i} & $5720.43$ & 21, RedL & 3.0 & $5720.44\pm0.02$ & $\phantom{1}\phantom{1}0.09\pm1.23$ & $0.25\pm0.05$ & $0.45\pm0.02$ \\
Cr~\textsc{i} & $5991.07$ & \phantom{1}3, RedU & 4.3 & $5991.05\pm0.01$ & $\,\,-7.61\pm2.11$ & $0.13\pm0.03$ & $0.23\pm0.01$ \\
Fe~\textsc{i} & $6083.66$ & 5, RedU & 3.3 & $6083.42\pm0.02$ & $-11.71\pm1.14$ & $0.24\pm0.05$ & $0.43\pm0.02$ \\
Fe~\textsc{i} & $6636.96$ & 13, RedU & 4.8 & $6636.51\pm0.02$ & $-20.36\pm0.77$ & $0.21\pm0.04$ & $0.37\pm0.01$ \\
\Hei & $6678.15$ & 14, RedU & 3.2 & $6677.93\pm0.03$ & $\,\,-9.75\pm1.45$ & $0.50\pm0.08$ & $0.91\pm0.02$ \\
Fe~\textsc{i} & $6703.57$ & 14, RedU & 3.1 & $6702.99\pm0.04$ & $-25.69\pm1.80$ & $0.54\pm0.09$ & $0.99\pm0.03$ \\\hline
\end{tabular}
\tablefoot{\\ \tablefoottext{a}{Line list is taken from the 
NIST Atomic Spectra Database \citep{NIST_ASD}.}\\\tablefoottext{b}{  Ca~\textsc{ii}~H is detected in two consecutive orders 25 and 24 of the BLUE arm, with much lower S/N for order 24, so only the detection in order 25 is indicated here.}}
\label{aptab3}
\end{table*}

\section{Fitting atmospheric models to 2M1115 SED}\label{sed}
\begin{figure}
    \centering
    \includegraphics[width=1.0\linewidth, trim = {0.9cm 0.4cm 2cm 0.5cm}, clip]{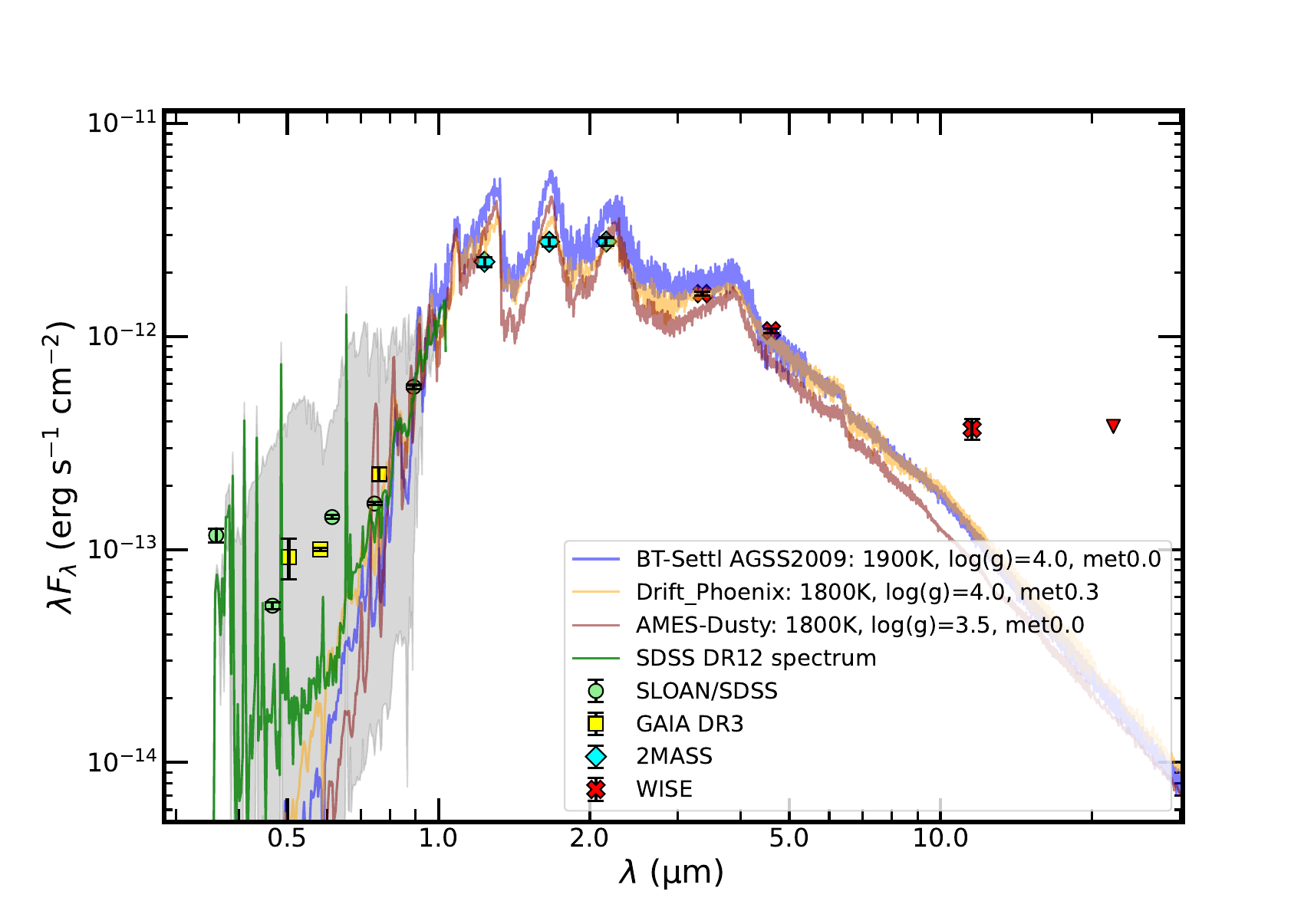}
    \caption{Least-$\chi^2$ fits to the SED of 2M1115 with DRIFT-PHOENIX, BT-Settl-AGSS2009 and AMES-Dusty atmospheric models. The overall best fit with the minimum $\chi^2$
    corresponds to DRIFT-PHOENIX model at $\Teff=1800\pm50$~K, log(\textsl{g})$=4.0\pm0.25$ and metallicity$=0.3\pm0.15$. The SDSS DR12 spectrum for 2M1115 is also shown, along with its flux uncertainty (grey shaded region). Note that the flux in the WISE $W4$ band is a 2$\sigma$ upper limit, denoted by the red downward-pointing arrow, and is not included in the fitting process. The unWISE catalogue (\citealp{lang16}; see \url{https://unwise.me/photsearch/}) however reports a W4 detection consistent with this upper limit.}
    \label{apfig5}
\end{figure}

Using VOSA, we queried the existing SDSS (\textit{u, g, r, i, z}), \textit{Gaia} DR3 (\textit{G}, $G_{\mathrm{BP}}$, $G_{\mathrm{RP}}$), 2MASS (\textit{J, H, K$_S$}) and WISE (\textit{W1, W2, W3, W4}) photometry of 2M1115. Since $W4$ magnitude is a lower limit, we exclude it from the analysis and fit the remaining 14 photometry points (refer Table~\ref{aptab5}) with BT-Settl AGSS2009 \citep{allard2012, allard2013}, AMES-Dusty \citep{chabrier2000, allard2001} and DRIFT-PHOENIX \citep{phoenix2003, phoenix2008, phoenix2011} atmospheric models constrained to near-solar metallicities ($-0.5$ to $0.5$), $\log(g)$ range 3--4 \citep[based on the spectral type L2$\gamma$;][]{kirkpatrick2005, theissen2018} and a temperature range of 1200--2500~K \citep[based on existing \Teff~estimate of $1724^{184}_{-38}$~K; ][]{theissen2018}. The extinction is currently unknown for the object; but it is still well within the Local Bubble given its location and hence extinction is likely negligible. So we assumed $A_V=0$ for our analysis. We do note, however, that the presence of an accretion disk around the target can lead to some intrinsic extinction within the system; this extinction is currently not constrained due to lack of sufficient data, and has not been accounted for in our analysis. Figure~\ref{apfig5} shows the best-fitting model from each of the three model grids that gives the minimum  $\chi^2$ with the observed photometry. From the Figure, we see a clear near-UV and optical flux excess compared to the model as well as MIR excess prominent towards the W3 band (indicative of a disk around the object), as previously noted in \cite{theissen2018}. The overall best-fit with least $\chi^2$ that fits 13 out of the 14 data points (with IR excess from $W3$) corresponds to DRIFT-PHOENIX models at \Teff$=1800\pm50$~K, $\log(g)=4\pm0.25$ and metallicity of $0.3\pm0.15$. VOSA also predicts the object parameters corresponding the least $\chi^2$ based on a polynomial fit to the $\chi^2$ vs. parameter-value space for each of the model grids used. Additionally, for each model grid, statistical analysis of the top least $\chi^2$ fits also gives average values of the best-fitting model parameters. To arrive at the optimal value of the target parameters, we took the set of parameters corresponding to the least $\chi^2$ fits, those predicted from the above-mentioned polynomial fit (as long as the value corresponding to the minimum $\chi^2$ falls within the constrained parameter range for the fit) as well as those derived from the average of the best-fits, for each of the three model grids. The rms of these sets of values give the following parameters for the target, which we adopt for our analysis: $\Teff=1816\pm63$~K, $\log(g)=3.83\pm0.24$, metallicity $=0.24\pm0.13$ and the integrated flux $F_{\mathrm{tot}}=(3.68\pm0.28)\times10^{-12}$~\fluxinteg. We derived a bolometric luminosity of $\Lbol=(2.34\pm0.29)\times10^{-4}\,\lsun$ based on the obtained $F_{\mathrm{tot}}$ and the \textit{Gaia} distance for the target. This \Lbol together with the obtained \Teff above gives a radius of $\Rp=1.54\pm0.14\,\rj$. As a cross-check on the VOSA results, we also fit the model grids to the SED of 2M1115 by including its SDSS DR12 spectrum (green curve in Figure~\ref{apfig5}) instead of the SDSS photometry. The resulting $\chi^2$ gives best-fitting models consistent with those from VOSA, adding robustness to the results derived above.

\begin{table}
\centering
\caption{Photometry for 2M1115 used for the SED fit.}
\begin{tabular}{c c c}
\hline\hline
Source & Filter & Brightness (mag) \\
\hline 
   SDSS & $u$ & $m_\mathrm{u}=21.06\pm0.08$ \\
        & $g$ & $m_\mathrm{g}=21.56\pm0.04$ \\
        & $r$ & $m_\mathrm{r}=20.23\pm0.02$ \\
        & $i$ & $m_\mathrm{i}=19.86\pm0.02$ \\
        & $z$ & $m_\mathrm{z}=18.30\pm0.02$ \\
   \textit{Gaia} & $G$ & $M_G=17.13\pm0.11$ \\
   2MASS & $J$ & $M_J=12.31\pm0.12$ \\
        & $H$ & $M_H=11.29\pm0.12$ \\
        & $K_S$ & $M_{K_S}=10.52\pm0.12$ \\
   ALLWISE & $W1$ & $M_{W1}=9.81\pm0.11$ \\
        & $W2$ & $M_{W2}=9.27\pm0.11$ \\
        & $W3$ & $M_{W3}=7.49\pm0.16$ \\
\hline                                   
\end{tabular}
\tablefoot{For the apparent 2MASS, \textit{Gaia} and WISE magnitudes, see Table~\ref{tab1}. The $u$, $g$, $r$, $i$, $z$ magnitudes are from SDSS DR12 \citep{sdss12} and the $W1$, $W2$, $W3$ magnitudes are from the ALLWISE catalogue \citep{wise2014}. The absolute magnitudes for the 2MASS, \textit{Gaia} and WISE bands have been computed based on the parallax-based distance estimate of $45.21\pm2.20$~pc. $M_J$, $M_H$ and $M_{K_S}$ are $\sim1.2\sigma$ brighter than the previously reported values in \cite{theissen2018}.} 
\label{aptab5} 
\end{table}

\section{Mass accretion rates from SDSS epochs} \label{sdss}
We downloaded the existing, public, moderate-resolution optical spectrum for 2M1115 from the two SDSS epochs DR9 \citep{sdss9} and DR12 \citep{sdss12} using the SDSS Science Archive Server (SAS). For the detected \Hi emission lines H3--H7 in the spectrum, we estimated a local continuum as the median flux within $\pm40$~\AA\ around the line, devoid of $\pm10$~\AA\ region centred on the line. After subtracting the continuum from the spectrum, we fit a single Gaussian profile to the line and estimated the line flux and line luminosities. The uncertainty in the line fluxes are estimated as the rms of the corresponding local continuum. We then used the planetary scaling relations from \cite{aoyama2021} to estimate accretion luminosities and mass accretion rates for the object, assuming the same values for \Mp and \Rp as in Section~\ref{mar}. The resulting line parameters for the lines H3--H7 are given in Table~\ref{aptab4}. The average accretion luminosity $\log(\Lacc/\lsun)$ calculated from lines H3--H7 is $-4.68\pm0.14$ for the DR9 epoch and $-4.30\pm0.14$ for the DR12 epoch. The corresponding mass accretion rates $\log(\mdot/\mjyr)$ in the two epochs are $-8.00\pm0.32$ and $-7.62\pm0.32$ respectively. In the DR9 epoch, the mass accretion rate estimated from \Ha seems to be slightly higher than those estimated from H4 and H5 but is consistent with those from H6 and H7 within uncertainties. In the DR12 epoch, the individual mass accretion rates derived from all the \Hi lines are consistent with each other within the uncertainties, except for that from H7 which is slightly higher. 

\renewcommand{\arraystretch}{1.2}
\begin{table}
\centering
\caption{Line fluxes, luminosities and mass accretion rates for 2M1115 from SDSS DR9 and DR12.}
\scalebox{1}{
\[
\begin{array}{llll}
\hline \hline
\multicolumn{1}{c}{\text{Line}} & \multicolumn{1}{c}{\text{Parameter}} & \multicolumn{1}{c}{\text{SDSS DR9}} & \multicolumn{1}{c}{\text{SDSS DR12}} \\
\hline
\Ha & F\mathrm{_{line}/10^{-15}\,[\fluxinteg]} & \phantom{1}\,7.28\pm0.03 & \phantom{1}\,6.78\pm0.02 \\
 & \logll & -6.33\pm0.04 & -6.36\pm0.04\\
 & \logla & -4.40\pm0.12 & -4.43\pm0.12 \\
 & \logm & -7.72\pm0.31 & -7.75\pm0.31 \\
\hline  
\Hb & F\mathrm{_{line}/10^{-15}\,[\fluxinteg]} & \phantom{1}\,0.68\pm0.03 & \phantom{1}\,2.22\pm0.03 \\
 & \logll & -7.36\pm0.05 & -6.85\pm0.04\\
 & \logla & -4.93\pm0.13 & -4.49\pm0.12 \\
 & \logm & -8.25\pm0.32 & -7.81\pm0.31 \\
\hline 
\mathrm{H\gamma} & F\mathrm{_{line}/10^{-15}\,[\fluxinteg]} & \phantom{1}\,0.41\pm0.03 & \phantom{1}\,1.64\pm0.05 \\
 & \logll & -7.58\pm0.05 & -6.98\pm0.04 \\
 & \logla & -4.84\pm0.15 & -4.33\pm0.14 \\
 & \logm & -8.16\pm0.32 & -7.65\pm0.32 \\
\hline 
\mathrm{H6} & F\mathrm{_{line}/10^{-15}\,[\fluxinteg]} & \phantom{1}\,0.29\pm0.04 & \phantom{1}\,1.29\pm0.05 \\
 & \logll & -7.73\pm0.07 & -7.08\pm0.05\\
 & \logla & -4.72\pm0.16 & -4.18\pm0.16 \\
 & \logm & -8.04\pm0.33 & -7.50\pm0.33 \\
\hline  
\mathrm{H7} & F\mathrm{_{line}/10^{-15}\,[\fluxinteg]} & \phantom{1}\,0.27\pm0.04 & \phantom{1}\,1.02\pm0.05 \\
 & \logll & -7.76\pm0.08 & -7.19\pm0.05\\
 & \logla & -4.53\pm0.16 & -4.06\pm0.16 \\
 & \logm & -7.85\pm0.33 & -7.38\pm0.33 \\
\hline  
\end{array}
\]
}
\tablefoot{Calculation of line fluxes are based on Gaussian fits to the respective lines. Accretion luminosities are calculated from line luminosities using \cite{aoyama2021} scaling relations.}
\label{aptab4}
\end{table}

\end{appendix}

\end{document}